\documentclass[12pt,preprint]{aastex}  
\usepackage{amssymb,amsmath} 
\usepackage{geometry}                        
\usepackage[parfill]{parskip}  
\usepackage{graphicx}
\usepackage{natbib}
\usepackage{amssymb}
\usepackage{epstopdf}
\usepackage[T1]{fontenc}
\usepackage[latin9]{inputenc}
\def\farcs{\hbox{$.\!\!^{\prime\prime}$}}
\def\fs{\hbox{$.\!\!^{\rm s}$}}

\begin{document}

\title{Multi-frequency Radio Measurements of \\SN 1987A over 22 Years}

\author{Giovanna Zanardo\altaffilmark{1}, L. Staveley-Smith\altaffilmark{1,a}, Lewis Ball\altaffilmark{2}, B. M. Gaensler\altaffilmark{3,b}, M. J. Kesteven\altaffilmark{2}, R. N. Manchester\altaffilmark{2,c}, C. -Y. Ng\altaffilmark{3}, A. K. Tzioumis\altaffilmark{2} and T. M. Potter\altaffilmark{1}} 

\altaffiltext{1} {International Centre for Radio Astronomy Research (ICRAR), School of Physics M013, The University of Western Australia, Crawley, WA 6009, Australia.}
\altaffiltext{2} {CSIRO Australia Telescope National Facility, PO Box 76, Epping, NSW 1710, Australia.}
\altaffiltext{3}{ Sydney Institute for Astronomy (SIfA), School of Physics, The University of Sydney, NSW 2006, Australia.}
\altaffiltext{a} {Premier's Fellow in Radio Astronomy}
\altaffiltext{b}{Australian Research Council Federation Fellow}
\altaffiltext{c}{CSIRO Fellow}
\email{giovanna.zanardo@physics.uwa.edu.au}
\email{}

\begin{abstract}
We present extensive observations of the radio emission from the remnant of SN 1987A made with the Australia Telescope Compact Array (ATCA), since the first detection of the remnant in 1990. The radio emission has evolved in time providing unique information on the interaction of the supernova shock with the circumstellar medium. We particularly focus on the monitoring observations at 1.4, 2.4, 4.8 and 8.6~GHz, which have been made at intervals of 4 -- 6 weeks. The flux density data show that the remnant brightness is now increasing exponentially, while the radio spectrum is flattening. The current spectral index value of $-0.68$ represents an $18\pm3\%$ increase over the last 8 years. The exponential trend in the flux is also found in the ATCA imaging observations at 9~GHz, which have been made since 1992, approximately twice a year, as well as in the 843 MHz data set from the Molonglo Observatory Synthesis Telescope from 1987 to March 2007. Comparisons with data at different wavelengths (X-ray, H$\alpha$) are made. The rich data set that has been assembled in the last 22 years forms a basis for a better understanding of the evolution of the supernova remnant.

\end{abstract}


\section{Introduction}

Supernova (SN) 1987A in the Large Magellanic Cloud, as the first naked-eye supernova to occur since the invention of the telescope, has become one of the most studied objects outside the solar system. Immediately after its discovery was announced on 23 February 1987, most telescopes in the southern hemisphere started observing its evolution. 
In addition to electromagnetic radiation, a burst of neutrinos was detected from the SN. The 25 neutrinos simultaneously detected by KamiokaNDE II in Japan, the Irvine-Michigan-Brookhaven detector in Ohio and the Baksan Neutrino Observatory in Russia, were more than enough to confirm the theoretical predictions made for the core collapse of a massive star \citep{arn89}. The detection of neutrinos allowed astronomers to {\em view} the collapse of the progenitor: Sanduleak (Sk) $-69^{\circ}\;202$. Records of Sk $-69^{\circ}\;202$ indicated that it had a mass close to 20 $M_{\odot}$ and, for 90 per cent of its life, was a blue star (Panagia, 2000). In the last million years of its life time, the star turned into a red supergiant and, following an evolution which is still somewhat uncertain, about 20,000 years before the explosion  returned towards the blue (Woosley \& Ensman 1988;  Crotts \& Heathcote 2000). 

SN 1987A has enabled the observation of a peculiar class of Type II events at close proximity. Its circumstellar structure was revealed in its intrinsic beauty by the Hubble Space Telescope. Observations showed a triple-ringed structure surrounding and illuminated by the SN (e.g. Burrows et al. 1995). The two outer rings appear to be the caps of an hour-glass shaped structure enveloping the SN itself.  The inner ring, also referred to as the equatorial ring, is believed to represent an equatorial density enhancement in the circumstellar medium (CSM), located at the interface between a dense wind emitted from an earlier red-giant phase of the progenitor star and a faster wind emitted by this star in more recent times (Crotts, Kunkel \& Heathcote 1995, Plait et al. 1995). The equatorial ring is the remnant of the stellar mass loss of the progenitor star. It is still under debate why the CSM remnant became concentrated into a ring rather than distributed spherically.
To explain the ring construction, some recent theories surmise that the progenitor of SN 1987A was  actually a merger (e.g. Morris \& Podsiadlowski 2007).

An initial radio outburst from SN 1987A peaked on day 4 and then decayed rapidly (Turtle et al. 1987). 
The radio flux of SN 1987A was found to rise again from noise in mid-1990, around 1200 days after the explosion, by the Molonglo Observatory Synthesis Telescope (MOST) (Turtle et al. 1990) and then by the Australia Telescope Compact Array (ATCA) (Staveley-Smith et al. 1992), signifying the birth of a radio remnant. Since the remnant detection, the radio emission of SN 1987A has shown a steady increase, surpassing the radio brightness of the initial prompt emission phase (Manchester et al. 2002, and references therein). The emission is generally thought to be synchrotron emission from electrons accelerated by expanding shock waves associated with the supernova explosion (Ball \& Kirk 1992). Accounting for uncertainties in both radio and optical measurements, Ng et al. (2008) suggested that the radio emission could have reached the optical inner ring around day ~6000-6500. This would justify the increase in the radio luminosity as well as match predictions made by Chevalier \& Dwarkadas (1995), Gaensler et al. (1997) and Manchester et al. (2002). Contemporaneously, increase in emission has been observed in X-ray wavelengths, firstly by the late Roentgen Satellite (ROSAT) which detected a slow but steady increase in flux over the early years (Gorenstein, Hughes \& Tucker 1994; Hasinger et al. 1996). From day $\sim$4600, subsequent observations with the {\it Chandra} X-ray Observatory have shown that the  brightening in soft X-rays has been characterised by an exponential trend, with a first upturn between days ~3500-4000  (Burrows et al. 2000), and a second one between days ~6000-6500, emphasizing the time when the blast wave presumably made contact with the dense protrusions of the CSM (Park et al. 2005). 

In more recent years, the SN ejecta themselves have become more difficult to observe in the optical, due to their faintness and the increasing brightness of the equatorial ring. The ring appears to have evolved into a clumpier structure, that now resembles a {\it pearl necklace}, with small regions of enhanced H$\alpha$ emission, or {\it hot spots} (Lawrence et al. 2000), appearing just inside the equatorial ring. Since year 2000, analysis of the super-resolved ATCA images at 9~GHz (resolution $\sim 0.5''$) have been showing a limb-brightened shell morphology. At this frequency, according to recent analysis of the shell structure, the eastern lobe is, consistently at all epochs, $\sim$ 40\% brighter than its western counterpart (Ng et al. 2008). In ATCA observations at 36 GHz made on days 7897-7902, the measured flux density ratio between the two lobes is similarly $\sim$ 1.3 (Potter et al. 2009). Based on 9~GHz images, it also may be that the eastern lobe is located farther from the SN site than the western one, suggesting an asymmetry in the initial expansion of the SN ejecta (Gaensler et al. 2007). These features of the remnant structure can be seen in recent multi-wavelength images (see Fig. 9 in Potter et al. 2009). The current remnant size, as obtained from data at 36 GHz, is $1''.70\pm0''.12$ (Potter et al. 2009). 

This paper presents the ATCA database of radio observations after 22 years of observations of SN 1987A, specifically at 1.4, 2.4, 4.8 and 8.6 GHz, as recorded from mid-1990 to February 2009.  In {\S~2} we first describe the procedure for imaging the ATCA data. In particular, we discuss improvements to the derivation of the flux densities at the higher frequencies. In {\S~3} we present ATCA flux densities in conjunction with those measured by MOST (Ball et al. 2001), and we present the related spectral indices. In {\S~4}, we discuss the current situation of the radio remnant. We compare the results in the radio to observations in the X-ray, H$\alpha$ and infrared (IR) wavelengths. We analyse the evolution of the SN and its environment in light of past predictions and existing models of SN 1987A. Previously published ATCA data at 1.4, 2.4, 4.8 and 8.6 GHz until 2001 February 2 (Manchester et al. 2002), and the new data are, for convenience, listed in their entirety in this paper.

\section{Observations}

\subsection{ATCA campaign}

Two types of observations have been made of SN 1987A using the ATCA: monitoring and imaging. Since its detection on Day 1200, the developing remnant has been monitored with various array configurations at 1.4, 2.4, 4.8 and 8.6 GHz. The monitoring observations have been made regularly every 1 -- 2 months (Manchester et al. 2002; Staveley-Smith et al. 2007). 

By Day 1786, the remnant became sufficiently bright to be imaged at a wavelength of  3 cm (Staveley-Smith et al., 1993), giving the opportunity to study its expansion (Gaensler et al. 1997). Since then, the source has been observed every six months with 6 km array configurations at 9 GHz with longer integration times ($\sim$12hr) (Gaensler et al. 2007; Ng et al. 2008).  As a terminology note, even though these full-synthesis observations of the source have been performed simultaneously at two frequencies in the range 8.1 to 9.1 GHz,  the flux densities derived from the images are denoted herein, both in the following text and figures, as derived from {\it 9 GHz} observations.

\subsection {Monitoring}

Flux density monitoring observations of the SN remnant are made simultaneously at two pairs of frequencies, either 1380 and 2496 MHz (2368 MHz before mid-1997) or 4790 and 8640 MHz. A 128 MHz bandwidth is used at all frequencies. The two frequency pairs are observed alternately, with 3 min on phase calibrators before and after pointing at the source for 20 min time intervals. The duration of the observation has been varied over the years, depending on the brightness of the source and, for the last three years, has been $\sim$6 hrs per epoch. Observations are centred approximately $10''$ south of the SN 1987A position [RA $05^{\rm h}\;35^{\rm m}\;28\fs03$, Dec $-69^{\circ}\;16'\;11\farcs79$ (J2000)] (Reynolds et al. 1995). The primary calibrator, used to set the flux-density scale, is PKS~B1934-638.  PKS~B1934-638 is assumed to have flux densities of 14.95, 11.14, 5.83 and 2.84 Jy at 1.4, 2.4, 4.8 and 8.6 GHz, respectively. The phase calibrators are PKS~B0530--727, PKS~B0407--658 and (at 4790/8640 MHz) PKS~B0454-810. In conditions of poor phase stability,  at the higher frequencies, PKS~B0454-810 is used as an alternative to PKS B0530--727.

\subsection {Data reduction}

The ATCA measures complex fringe visibilities in the $u-v$ plane (Frater, Brooks \& Whiteoak 1992). Data are reduced in {\sc miriad}\footnote{http://www.atnf.csiro.au/computing/software/miriad/}.  Visibility data are edited and calibrated in both phase and amplitude.  Images are formed using baselines longer than 3k$\lambda$, since shorter baselines are often contaminated by emission from strong sources outside the primary beam (primarily 30 Doradus). 

It is known that self-calibration (Pearson \& Readhead 1984; Cornwell \& Fomalont 1989), applied to strong (flux peak $S$ > 100 mJy/beam) and compact (area covered by source < 15 times the area of the synthesised beam) sources, often improves the image without the need for further editing (Burgess \& Hunstead 1995). To improve the quality of the SN 1987A images at 4.8 and 8.6 GHz and, thus, to reduce the error in deriving the flux density, the self-calibration procedure {\sc selfcal} has been incorporated in the reduction process. The self-calibration solution interval was set to 5 minutes. The preliminary {\sc clean} model is constructed by using a small number of iterations, whilst a deeper {\sc clean}ing was performed after phase self-calibration.

Self-calibration has been applied to all SN 1987A datasets recorded from 2007, when the source's flux density reached 80 mJy at 8.6 GHz. As a result of the process, the self-calibrated images are of improved quality and are often characterized by a flux density higher than those normally calibrated. Given the remarkably good images produced from even unpromising data, older datasets have been retrospectively re-processed with the incorporation of self-calibration,  whenever the application of such procedure has resulted in an improved image. We note that self-calibration has enhanced the ATCA images derived from datasets dated back to Day 5729. Light curves for normal and self-calibrated data are compared in Figure 1, while one example  of the differences between the normal and self-calibrated images is given in Figure 2. 

As regards observations at 1.4 and 2.4 GHz, images back to Day 5729 have been {\sc clean}ed more deeply than previous observations.

\section{Results}

\subsection{Flux densities}

Monitoring data for SN 1987A at 1.4, 2.4, 4.8 and 8.6 GHz are listed in Table 1 and shown in Figures 3 and 4. Two uncertainties are associated with each measurement: a random error, representative of the noise in the image and the uncertainty in the fitting process, and a systematic error estimated by examining the scatter in the flux density of a nearby ($\sim$4') unresolved source J0536--6919 (Gaensler et al. 1997). Except at 8.6 GHz, the systematic errors are estimated from the scatter in the measured flux densities of J0536-6918 since day 4000. As indicated by Manchester et al. (2002), at 8.6 GHz, J0536-6918 is outside the half-power radius of the primary beam, and systematic errors are taken to be 1.25 times the systematic errors measured at the 4.8 GHz. At the lower frequencies and at later times, the errors in the flux density estimations are dominated by the systematic errors.

The flux density was measured by integrating a Gaussian fit to the emission region at 1.4 and 2.4 GHz, while at 4.8 and 8.6 GHz the flux density was integrated over a polygonal region around the source without fitting. In Figures 3 and 4, ATCA flux densities derived from monitoring observations at 1.4, 2.4, 4.8, 8.6 and imaging observations at 9 GHz are plotted in the same graph with MOST data \footnote{http://www.physics.usyd.edu.au/ioa/wiki/index.php/Main/SN1987A} at 843 MHz (Ball et al. 2001).  The flux densities from the ATCA monitoring data at 8.6 GHz and the imaging observations at 9 GHz are in very close agreement, confirming the accuracy of the model-fitting technique used to analyse the latter.

The plots, both in linear and logarithmic scales,  show that there is a systematic trend in the flux increase at all frequencies. In particular, from day $\sim$3000, all ATCA observations appear to consistently follow an exponential trend. In Figure 4, the flux at each frequency is fitted to an exponential model. The fit parameters related to exponential fits from Day 5000 are reported in Table 2.  As regards MOST data, even though the related observations are not as regular as the ATCA ones, there is evidence of a steady increase that can also be fitted by an exponential curve from day $\sim$3000 onwards. MOST fitting parameters from Day 5000 are also listed in Table 2.  By looking at the overall development of the data at all frequencies, it has to be highlighted  that the resulting best fits significantly depart from both the simple linear fit proposed by Ball et al. (2001) and the quadratic trend proposed by Manchester et al. (2002), for flux densities measured up to day $\sim$5000. However, it is noted that both Ball et al. (2001) and Manchester et al. (2002) fitted the data within the limited time interval between Day 3000 and 5000, while the exponential curve appears to closely match the data only after day $\sim$4000. 

To further quantify the flux density increase, the increase factor per year has been calculated since the discovery of the remnant, i.e. from day $\sim$1200. The increase rate per year of the flux density, $S$, is calculated as $(S(t+\tau)-S(t))/S(t)$, with $\tau$ approximately equal to 365 days. Local interpolation of the flux density values was performed when the observation dates did not match the yearly time intervals. The error on the rate values was derived from the errors affecting the two flux density values considered at the beginning and at the end of the yearly intervals. A graphical representation of the derived rates of flux density increase per year is shown in Figure 5. It emerges that the flux density at the higher frequencies has undergone, in general, a higher increase rate. This is corroborated by the decrease of $e$-folding time, $\Delta$, at higher frequencies (Table 2). As regards the increase rates in recent years, at $\sim$day 7070, i.e. Year 20 since the supernova explosion, the percent increase of flux density at 1.4, 2.4, 4.8 and 8.6 GHz has been of 16$\pm4$\%,17$\pm4$\%, 14$\pm5$\% and 18$\pm6$\%, respectively. While in Year 21, the flux density increase rates, given in the aforementioned frequency order, are 13$\pm3$\%,17$\pm4$\%, 12$\pm5$\% and 16$\pm6$\%. In Year 22, since the latest observations were made in 2009 February 1, and the yearly intervals run from June, the rates of increase were calculated over a time frame that is $\sim$40\% shorter than a year. However, it is worth noting that the resultant increase rates have a slightly smaller mean value of 11$\pm4$\%. 

\subsection{Spectral indices}

In Figure 6, values of the spectral index $\alpha$, where the flux density $S=\nu^{\alpha}$ and $\nu$ is the frequency, found by linear regression, are presented with yearly regularity since the discovery of the remnant. In this figure, from Day 1517 to Day 7084 the spectral index is calculated over five frequencies, i.e. MOST frequency at 843 MHz and the  ATCA frequencies at 1.4, 2.4, 4.8 and 8.6 GHz, while afterwards it is calculated over ATCA frequencies only.  Quoted errors are 1$\sigma$. It should be noted that there are some minor differences between the spectral index obtained by fitting flux densities from observations between 843 MHz and 8.6 GHz cm ($\alpha_{1}$) and from the four ATCA-only observations between 1.4 and 8.6 GHz ($\alpha_{2}$). These differences, calculated at yearly-spaced epochs, are highlighted in Table 3. The maximum deviation, $\Delta\alpha$, between the central values of the resultant $\alpha_{1}$ and $\alpha_{2}$ at a given epoch, is 0.057, while the average deviation is 0.025. This corresponds to an average percent difference, centred on $\alpha_{2}$, around $3\%$. However, since MOST data are available to Day 7321, while ATCA data are available to Day 8014, all spectral index values reported in Figures 7--8, i.e. from the discovery of the remnant to date, for consistency, were derived from ATCA-only frequencies. 

Figure 7 shows that the spectral index has been more or less flattening since day $\sim$2500, and this trend can be emphasized by a linear fit  in the form of $\alpha(t)=\alpha_{0} +\beta_{0}\times(t-t_{0})/\Delta$, where $t$ is expressed in days (d), $t_{0}=5000$ d and $\Delta=365$ d, $\alpha_{0}=-0.825\pm0.005$ and $\beta_{0}=0.018\pm0.001$. The comparison of the linearly fitted spectral index $\alpha'$ corresponding to Day 5000, i.e. $\alpha'_{5000d}=-0.826\pm0.019$, with that corresponding to Day 8000,  i.e. $\alpha'_{8000d}=-0.673\pm0.012$, leads to a $18\pm3\%$ increase of the spectral index over the last 8 years. For completeness, in Figure 8,  the spectral indices obtained from ratios of both the higher frequencies (4.8 and 8.6 GHz) to the lower frequencies (1.4 and 2.4 GHz), are plotted in the same graph as the $\alpha_{2}$ values. While spectral indices derived from the ratio of close frequencies, e.g. 2.4 and 4.8 GHz, exhibit a wider scatter, all ratios appear to have progressively flattened since days $\sim$2500--3000.

\section{Discussion}

\subsection{SN-CSM interaction} 
The radio flux density of the remnant as a function of time  and frequency is a guide to the conditions at the shock front. More specifically,  while the analysis of the emission at different frequencies provides details of the particle acceleration mechanism that is associated with the shock propagation, the evolution of the flux density in time provides insight into the properties of the material into which the shock is expanding. 

The observed steady increase of the radio emission over time is rather unusual for a SN twenty years after the explosion (Panagia 2000). Such increase is normally only seen  in the very early optical thick phase of radio supernovae (Weiler et al. 2002). The fact that the radio emission has been observed to rise at an exponential rate indicates that the propagating blast wave, and the increasing shock volume associated with that, is interacting with an increasingly dense region of the CSM, likely to be associated with the equatorial ring. 

In the frequency domain, as shown by the radio spectra, the emission appears to have a non-thermal power-law form, which is the signature of the synchrotron process. This is usually attributed to particle acceleration from the first-order Fermi mechanism, a process in which particles scatter between the upstream and downstream regions of the forward shock to gain energy (for review see Jones \& Ellison 1991). The spectrum of the accelerated particles can be expressed as a function of the electron spectral index, $\gamma$, through the relationship $N(p)\propto p^{-\gamma}$, where $N$ is the particle density, $p$ is the particle scalar momentum and $\gamma = (\sigma_{s} +2)/(\sigma_{s} -1)$, with $\sigma_{s}$ the compression ratio of the subshock (Jones \& Ellison 1991). It should be noted that the front of the forward wave consists of an extended precursor and a viscous velocity discontinuity (subshock) of a local Mach number that is smaller than the total Mach number of the shock wave. According to the model developed by Berezhko \& Ksenofontov (2006), that fits SN 1987A data given in Manchester et al. (2002), the compression ratio of the subshock can be taken as  $\sigma_{s} \sim \sigma / 2$,  $\sigma$ being the compression ratio of the total shock. The radio spectral index can then be related to the electron spectral index as $\alpha=(1-\gamma)/2$. Taking $\alpha= -0.68\pm0.01$, as the lowest value derived from SN observations in recent epochs, the resulting  value of the subshock compression ratio is $\sigma_{s}=3.20\pm0.04$, which is lower than the classical value 4,  obtained from $\rho_2 / \rho_1=(\gamma_{g}+1)/(\gamma_{g} -1)$, with $\gamma_{g}=5/3$ the ratio of specific heats of a monatomic or fully ionized gas, while $\rho_1$ and $\rho_2$ are the density of the upstream and downstream interstellar gas, respectively. Considering the fitted spectral indices $\alpha'_{5000d}$ and $\alpha'_{8000d}$,  the comparison between the corresponding $\sigma_{s}$ values leads to a $14\pm3\%$ increase of the compression ratio over the last 8 years. 

In the case of a young supernova remnant, i.e. a steadily brightening remnant, such as that of SN 1987A, it is expected that either magnetic entanglement or small-scale spatial fluctuations should be present (Ryutov et al. 1999). In this scenario, from a macroscopic point of view, the magnetic field can be characterized in terms of magnetic field density, $B$, which can be related to the compression ratio to estimate the magnetic discontinuity associated with the propagating blast wave. In the particular scenario of SN 1987A, where the forward shock has likely reached the higher density matter in the equatorial ring, the major field discontinuity is likely to be located by the forward shock, where the  pressure on the denser CSM in the ring will be much higher than the pressure on the CSM shocked by the reverse shock. It follows that the ratio of  the downstream to the upstream magnetic field densities can be assumed to be $B_{2}/B_{1}\propto \sigma$ (Berezhko \& Ksenofontov 2006), where, in recent epochs,  $\sigma\sim6.4$. Then, since the particle acceleration process becomes efficient at the discontinuity in the magnetic field density, the majority of the synchrotron emitting electrons is likely to be generated along the propagating forward shock. As a direct consequence of the increased compression ratio, the density of the particles accelerated by the forward wave has also increased. In particular, considering the fitted spectral index values $\alpha'_{5000d}$ and $\alpha'_{8000d}$, the particle density has changed from $N(p)\propto p^{-2.65\pm0.04}$ to $N(p)\propto p^{-2.35\pm0.02}$ in the last 8 years, thus underlying the fact that the acceleration process is becoming more efficient. The enhanced efficiency of particle acceleration should allow the forward wave to match the energy requirements for the production of cosmic rays (CRs). In turn, in presence of efficient CR acceleration, the magnetic field can grow via a non-resonant instability and  become orders of magnitude larger than that of the typical interstellar field (Bell 2004).

Detailed modeling of the hydrodynamic effects of shock propagation, which is beyond the scope of this paper, is needed to explain the exponential trend of the radio emission. However, previous models of SN 1987A that predicted to some extent the observed evolution in the radio include that of Berezhko \& Ksenofontov (2000, 2006), who developed a model which couples the CR acceleration process with the hydrodynamics of the thermal gas (first proposed by Duffy, Ball \& Kirk 1995). Based on the flux density data reported by Gaensler et al. (1997) from Day 918 to Day 3325, Berezhko \& Ksenofontov (2000) (BK00) provide theoretical predictions up to Day 4780. In Figure 9, the observations at 1.4 GHz are compared to the BK00 model curve. While the BK00 predictions show, up to Day 4780, a continuous increase of the radio emission, after day $\sim3000$ the increase rate appears to be somewhat lower than what has been observed. In Figure 9, the observations at 1.4 GHz are also compared to the flux densities derived from the model presented in Berezhko \& Ksenofontov (2006) (BK06) that corresponds to the upstream magnetic field $B_{1}\sim$ 3 mG and  $\sigma\sim 3$.  The BK06 model is calibrated on the radio data given by Manchester et al. (2002) for Days 1970, 3834 and 5093, and provides theoretical predictions for Day 7300 and Day 8249. This model curve appears to have an increase rate higher than what observed, with predicted flux density on Day 8249 around twice the value measured on Day 8014. The model-data comparison in terms of spectral index is shown in Figure 7. It can be seen that BK06 predict steepening of the radio spectrum rather than the progressive hardening observed since day $\sim$ 2500. Kirk, Duffy \& Gallant (1996) have shown that partial trapping of the particles by structures in the magnetic field, which results in a sub-diffusive transport of particles at the shock front, can significantly soften the spectrum. 

Following a different approach, Chevalier \& Dwarkadas (1995) (CD95) attribute the SN radio emission observed since mid 1990 to the shock interaction with an H{\sc ii} region located inside the optical ring. While the radio emission would be insignificant before the CSM ring collision because of the low-density surrounding medium, the CD95 model explains the time frame required for the turn-on of the radio emission and for the shock wave to reach the dense ring. In particular, CD95 derived that the H{\sc ii} region, specifically modeled as a spherical shell created by the blue supergiant progenitor star in the swept-up red supergiant wind, would have delayed the encounter of the blast wave with the equatorial ring to year $2005\pm3$. By then, a drastic increase in the radio flux was predicted. This timing matches that of the radio emission from SN 1987A as observed to date.

\subsection{SN 1987A at other wavelengths}

From theoretical expectations, the SN 1987A  remnant, in its current stage, can be conceived as a structure including the following regions: {\it a}) the unshocked ejecta in the centre; {\it b}) the ejecta shocked by the reverse shock; {\it c}) the CSM at the inner edge of the equatorial ring shocked by the reverse shock; {\it d}) the CSM at the outer edge of the equatorial ring shocked by the forward shock; {\it e}) the CSM within the inner ring radius which, after being shocked by the forward shock, have been shocked also by the reflected shock. The reflected shock would have appeared when the forward wave suddenly encountered the higher density CSM in the equatorial ring. Observations at optical/near-infrared wavelengths have helped to distinguish the emission sites $a-c$, while X-ray and radio observations have captured the different types of emission coming from the shocked CSM localized in sites $c-e$.

The X-ray, H$\alpha$ and IR fluxes  ares compared to the radio flux in Figure 10. X-ray data from {\it Chandra} {\sc ACIS} are updated to Day 7271 (Park at al. 2007). {\it ROSAT} data (up to Day 3950) and {\it XMM-Newton} data (Days 4712, 5156 and 5918) are taken from Haberl et al. (2006). H$\alpha$ flux values are taken from Sonneborn et al., (1998), Smith et al. (2005) and Heng et  al. (2006). IR data (Days 6190 and 7137) are as reported by Dwek et al. (2008). The radio flux is derived after integrating the flux density in the frequency range between 1 and 10 GHz. Given the scatter of the spectral index values obtained from observations at the four ATCA frequencies (see Figures 7--8), the linear fit values, $\alpha'$, corresponding to each epoch, were used to calculate the radio flux. 

As shown in Figure 10,  from early epochs to days 4500--5000,  the evolution of the radio flux seems to be matched by that of {\it ROSAT} X-ray data. It should be noted that, while early {\it ROSAT} data did not clearly differentiate between soft and hard components of the X-ray flux, the fact that the SN was initially detected as a faint soft X-ray ($\sim0.5-2$ keV) source has been discussed by Hasinger et al. (1996). On the other hand, in the {\it Chandra} data, soft and hard ($\sim3-10$ keV) X-ray components are clearly distinguished. From days 6000-6500 onwards, a similarity between the increase rate of the radio flux curve and that of the hard X-ray flux is evident.

Up to days 4500--5000, the soft X-ray light curve can be fitted by a power-law curve $F(t)\propto t^{2}$. The similarity between the increase rates of the soft X-ray and radio fluxes, within this part of the light curves, might imply that, up to days 4500--5000, the soft X-ray and radio emission occupied the same volumes. Around days 4500--5000, the soft X-ray flux started to rise at an exponential rate much steeper than that of the radio flux, i.e. $F\propto e^{t/\Delta}$, with $t$ expressed in days (d) and $\Delta = 875\pm3$ d. To justify such a change of the flux increase rate, Park et al. (2005) suggested that the soft X-rays could be produced by the decelerated forward shock entering the dense protrusions of the equatorial ring. In fact, since the departure from the initial power law curve can be identified around the time when hot spots were discovered on both sides of the ring (Lawrence 2000), such discontinuity in the soft X-ray light curve might flag the beginning of the interaction between the shock wave and the dense gas all around the equatorial ring. Accordingly, if the blast wave struck the ring, at this time a reflected shock must have been generated due to impact with the higher density matter. The reflected shock would then slow the X-ray emitting gas to the velocity of the transmitted shock.  This twice-shocked gas, characterised by much greater density and higher temperature than the gas behind the blast wave, would then be likely to have significantly increased soft X-ray emissivity. In such scenario, whose plausibility has been discussed by Zhekov et al. (2009), the exponential rise  observed in the soft X-ray flux since days 4500--5000  could be the result of the increasing volume of the shocked gas by the expanding forward shock, as well as of the twice-shocked gas. If this is the case, the soft X-ray overall emission would be now bounded by the forward and reverse shocks. 

According to Ng et al. (2008), the radio emission overtook the optical inner ring around days 6000-6500. Evidence that the inner ring is being swept by the shock wave can be found in the IR data. In particular, the IR data plotted in Figure 10 show that the IR flux has doubled from Day 6190 to Day 7137. The observation that the ratio of the IR to the soft X-ray flux  has decreased from  5 to 3 from the first to the second observation epoch, suggests that the propagation of the shock has been destroying the dust grains (Dwek et al. 2008).

As regards the hard X-ray flux, from days $\sim$5000--5500 to days $\sim$6000--6500, the flux curve looks somewhat similar to that outlined by the H$\alpha$ flux values at corresponding epochs, while, after days $\sim$6000--6500, as mentioned above, the rate of the flux increase seems to have slowed and have become closer to that of the radio flux. The similarity between the trend of hard X-ray and radio fluxes has been noted before (Park et al. 2005, 2007). Park et al. (2007) also noted that, since the morphology of hard X-ray images are no longer distinguishable from that of soft X-ray images, the origin of hard X-ray emission is uncertain. It is known that the H$\alpha$ emission essentially comes from the inner edge of the equatorial ring, as a result of the impact of the shock wave with the dense obstacles that protrude inward from the ring (Michael et al. 2000, 2003). However, since the data from Day 3743 to Day 6577 are  affected by large systematic uncertainties, it is difficult to discern between the two limiting behaviours of the H$\alpha$ emission, specifically $F(t)\propto t$ and $F(t)\propto t^{5}$ (Heng et al. 2006). Heng et al. (2006) theorised that the H$\alpha$ emission consists of  two main components, both related to the reverse shock: {\it a}) a fainter inner component, that comes from freely-streaming hydrogen atoms in the supernova debris and originate within the reverse shock emission; {\it b}) a brighter component, that comes from the surface of the reverse shock. The fact that, from days $\sim$5000--5500 to days $\sim$6000--6500, the magnitude of the H$\alpha$ and hard X-ray emissions seems to  align to some extent, could thus indicate that the hard X-ray emission might be associated with the reverse shock.

In Figure  11, the ratios of the X-ray, H$\alpha$, and IR to the radio flux are compared from Day 1448 to Day 7271. It can be noted that, while some measurements suffer from greater error, some features can be identified. In particular, the ratio of the soft X-ray to the radio flux is practically constant from day $\sim$1500 to day $\sim$ 4000. Then, from day $\sim$4500 to day $\sim$7000, the ratio increases by a factor of  6. The increase in flux ratio appears to have flattened after day $\sim$6600. Park et al. (2005) noted the fact that this can be associated with the drastic deceleration of the expansion velocity of the X-ray emitting hot gas, which was detected around day $\sim$6200. This could correspond to the time when the forward shock has started to decelerate, while overtaking the denser matter within the main body of the equatorial ring (Bouchet et al. 2006). Drastic deceleration in the remnant evolution is not confirmed by observations in the radio (Ng et al. 2008). As regards the ratios of H$\alpha$ to the radio flux, values have been varying from a low 0.6 on Day 3743 to unity on Day 5729, while appeared to have been decreasing afterwards. Ratios of IR to radio flux have increased by $\sim30\%$ from the first to the second epoch of observation. 
 
\subsection{SN 1987A and other radio supernovae}

The radio emission from Type II SNe can be understood with a model (Chevalier 1982) in which a synchrotron source generated by the blast wave interacts with an optically thick CSM, which becomes optically thin at larger radii. The density pattern of the CSM is presumed to be established by the nature of the progenitor and the mass-loss history associated with the progenitor wind. Long-term observations of radio SNe essentially allow us to probe the radial dependence of the surrounding CSM and to monitor the changes of the CSM  structure. Because of its near proximity, SN 1987A  is a Type II-P SN that was detected from the early stages, thus allowing detailed scrutiny of one of the most complex CSM surrounding a SN, as well as of the evolving  interaction between the blast wave and such CSM.

There are few Type II SNe detected from such early stages with well-sampled light curve (Weiler et al. 2002).
SN 1993J in M81 (Weiler et al. 2007) is a Type II-b SN that has been observed as early as one or two days after explosion. Due to the mass-loss of the progenitor before the explosion, SN 1993J has shown a radio evolution that is practically opposite to that of SN 1987A. More specifically, the radio emission of SN 1993J is characterized by an exponential decay, with a decline rate that is significantly steepening, and by a constant spectral index.  Such decay of the radio flux, which started to affect all observed frequencies  $\sim$1000 days after the SN discovery, has been interpreted in terms of a sudden decrease of the CSM density. Opposite is the case of SN 1996cr in the Circinus Galaxy (Bauer et al. 2008), a Type II-n SN strongly interacting with the CSM and observed in the radio within days of the explosion. The progenitor of SN 1996cr appeared to have changed evolutionary states just prior to the explosion. Thus,  similar to SN 1987A, SN 1996cr is potentially facing an increase of the remnant brightness in the future, as result of the possible interaction with a surrounding ring or a disk left from the progenitor explosion, which, however, has not been detected. 

As regards the small group of Type II-P SNe like SN 1987A, the more recent SN 2004dj in NGC 2403 (Stockdale et al. 2004a) and SN 2004et in NGC 6946 (Stockdale et al. 2004b) have been closely observed in the radio since their discovery. SN 2004dj,  detected in the radio only a few days after the explosion, is the brightest SN since SN 1987A and, to date, the only other Type II-P supernova  which has a well-sampled radio light curve. However, SN 2004dj is currently very faint in radio, as it is still in the transition from radio supernova to supernova remnant (Beswick et al. 2005). The emission structure of SN 2004et is an example of a Type II SN that could provide insight into the reasons behind the asymmetry of many SN remnants.  SN 2004et  is in fact characterized by an asymmetric shell that might be the result of anisotropies of the CSM or asymmetry in the CSM distribution (Mart{\'{\i}}-Vidal et al. 2007).

Examples of Type II SNe with well-sampled light curves, but observed as late as months or years after the explosion, are similarly aged Type II-L SN 1979C, observed since day $\sim$500 (Weiler et al. 1986; Montes et al. 2000), and SN 1980K, observed since day $\sim$100 (Weiler et al. 1986; Montes et al. 1998), as well as Type II-n  SN 1988Z, observed since $\sim$day 400 (Williams et al. 2002). Compared with SN 1987A, these SNe have shown a surrounding medium of less complex structure, as, by days $\sim$1500--2000 at the latest, their radio emission has declined at all frequencies. 

\section{Conclusions}

In this study, we have presented the results of 22 years of observations at radio wavelengths of SN 1987A.  In particular, we have discussed the evolution of the supernova remnant in light of the flux density derived from monitoring observations at 1.4, 2.4, 4.8, 8.6 GHz (ATCA), imaging observations at 9 GHz (ATCA), as well as from observations at 843 MHz (MOST).  

The fact that, at least since day $\sim$5000, the radio emission has been rising at an exponential rate, signifies that SN 1987A is undergoing a transition, i.e. the blast wave has reached a denser region of the CSM and is interacting with it. Moreover, the exponential increase rate of the flux density, which can be observed at  all frequencies,  is accompanied by a progressive flattening of the radio spectrum. A flatter radio spectrum corresponds to a flatter momentum spectrum for the particles accelerated by the shock front, which, in turn, is an indication of the increased production of non-thermal electrons and other cosmic rays. Therefore, the exponential increase of the radio light curve is likely to be a consequence of the increased  efficiency of the acceleration mechanism that generates synchrotron radiation at the forward shock.

Based on the analysis of the composite light curves, the blast wave is likely to have reached the inner protrusions of the inner optical ring  around days $\sim$4500--5000. Since then the soft X-ray flux has undergone an exponential increase much steeper than that measured at other  wavelengths. The increase of the soft X-ray flux appears to be the most outstanding consequence of the forward shock encountering the equatorial ring.  However, the radio emission has also been increasing exponentially, albeit at a slower $e$-folding rate.

At the moment, predictions on future developments of the radio remnant are still uncertain, primarily because of the poor understanding of the CSM surrounding the remnant, the insufficient knowledge of particle acceleration mechanisms, as well as of the evolution of the magnetic field. However, it is likely that the radio emission will further brighten, as the interaction of the shock wave with the dense CSM progresses.

\bigskip

\acknowledgments 
We would like to thank Robin Wark and Bjorn Emonts for their assistance during site-based ATCA observations, John Kirk and Brian Reville for useful discussions, Leonid Ksenofontov for providing the model data. The Australia Telescope Compact Array is part of the Australia Telescope, which is funded by the Commonwealth of Australia for operation as a National Facility managed by CSIRO. The Molonglo Observatory Synthesis Telescope is operated by the University of Sydney and supported in part by grants from the Australian Research Council. B. M. G. and C. -Y. N. acknowledge the support of Australian Research Council through grant FF0561298.

\clearpage


\clearpage


\begin{table}[htb]
\begin{center}
\caption{Flux densities measured at the four ATCA frequencies over the day range 918--8014. These are calibrated relative to the primary flux calibrator, PKS B1934-638. Day number is counted from UT 1987 February 23.}
\bigskip
\footnotesize
\begin{tabular}{cc|cc|cc|cc}
\tableline\tableline
\multicolumn{2}{c}{1.4 GHz} & \multicolumn{2}{c}{2.4 GHz} & 
\multicolumn{2}{c}{4.8 GHz} & \multicolumn{2}{c}{8.6 GHz} \\

\tableline
Day & Flux Density & Day & Flux Density & Day & Flux Density & Day & Flux Density \\
 	& (mJy) 	       &  	 & (mJy) 		&   	   & (mJy) 		  &  	    & (mJy) \\ 
\tableline

-	&		-		&	-	&		-		&	918	& $	0.0	\pm	0.3	$ &	-	&		-		\\
-	&		-		&	-	&		-		&	996	& $	0.0	\pm	0.3	$ &	-	&		-		\\
-	&		-		&	1199	& $	0.0	\pm	1.0	$ &	-	&		-		&	-	&		-		\\
1244	& $	0.0	\pm	1.2	$ &	-	&		-		&	1244	& $	0.0	\pm	0.3	$ &	-	&		-		\\
1271	& $	2.9	\pm	1.2	$ &		&		-		&	1270	& $	0.7	\pm	0.3	$ &	-	&		-		\\
-	&		-		&	-	&		-		&	1306	& $	1.2	\pm	0.3	$ &	-	&		-		\\
-	&		-		&	1315	& $	3.3	\pm	1.0	$ &	-	&		-		&	-	&		-		\\
-	&		-		&	-	&		-		&	1334	& $	1.5	\pm	0.3	$ &	-	&		-		\\
-	&		-		&	-	&		-		&	1366	& $	1.5	\pm	0.3	$ &	-	&		-		\\
1386	& $	4.9	\pm	1.2	$ &	1386	& $	3.5	\pm	1.0	$ &	1385	& $	2.0	\pm	0.3	$ &	-	&		-		\\
-	&		-		&	-	&		-		&	-	&		-		&	1388	& $	1.3	\pm	0.3	$ \\
-	&		-		&	-	&		-		&	1401	& $	2.2	\pm	0.3	$ &	-	&		-		\\
-	&		-		&	-	&		-		&	1402	& $	2.2	\pm	0.3	$ &	-	&		-		\\
-	&		-		&	-	&		-		&	1403	& $	2.3	\pm	0.3	$ &	-	&		-		\\
-	&		-		&	-	&		-		&	1404	& $	2.1	\pm	0.3	$ &	-	&		-		\\
-	&		-		&	-	&		-		&	1405	& $	2.1	\pm	0.3	$ &	-	&		-		\\
-	&		-		&	-	&		-		&	1407	& $	2.1	\pm	0.3	$ &	-	&		-		\\
-	&		-		&	-	&		-		&	1408	& $	2.6	\pm	0.3	$ &	-	&		-		\\
-	&		-		&	-	&		-		&	1409	& $	2.5	\pm	0.3	$ &	-	&		-		\\
-	&		-		&	-	&		-		&	1410	& $	2.7	\pm	0.3	$ &	-	&		-		\\
-	&		-		&	-	&		-		&	1431	& $	2.3	\pm	0.3	$ &	1432	& $	1.1	\pm	0.3	$ \\
-	&		-		&	-	&		-		&	1446	& $	2.6	\pm	0.3	$ &	-	&		-		\\
-	&		-		&	-	&		-		&	1460	& $	3.0	\pm	0.3	$ &	-	&		-		\\
1490	& $	7.6	\pm	1.2	$ &	-	&		-		&	-	&		-		&	-	&		-		\\
-	&		-		&	-	&		-		&	-	&		-		&	1501	& $	1.7	\pm	0.3	$ \\
-	&		-		&	-	&		-		&	-	&		-		&	1515	& $	2.3	\pm	0.3	$ \\
1518	& $	8.3	\pm	1.2	$ &	1518	& $	6.1	\pm	1.0	$ &	-	&		-		&	-	&		-		\\
1518	& $	9.6	\pm	1.2	$ &	1518	& $	6.2	\pm	1.0	$ &	1517	& $	2.8	\pm	0.3	$ &	-	&		-		\\
-	&		-		&	-	&		-		&	1525	& $	2.7	\pm	0.3	$ &	-	&		-		\\
-	&		-		&	-	&		-		&	-	&		-		&	1584	& $	2.6	\pm	0.3	$ \\
-	&		-		&	-	&		-		&	1587	& $	3.0	\pm	0.3	$ &	-	&		-		\\
1596	& $	10.2	\pm	1.2	$ &	1596	& $	5.6	\pm	1.0	$ &	1595	& $	4.5	\pm	0.3	$ &	1594	& $	2.2	\pm	0.3	$ \\
1637	& $	15.5	\pm	1.3	$ &	1637	& $	8.7	\pm	1.0	$ &	1636	& $	5.3	\pm	0.3	$ &	-	&		-		\\
1661	& $	15.5	\pm	1.3	$ &	1661	& $	9.2	\pm	1.0	$ &	-	&		-		&	1662	& $	2.5	\pm	0.3	$ \\
-	&		-		&	-	&		-		&	1663	& $	5.0	\pm	0.3	$ &	1663	& $	2.9	\pm	0.3	$ \\
-	&		-		&	-	&		-		&	1747	& $	6.0	\pm	0.4	$ &	-	&		-		\\

\tableline
\end{tabular}
\end{center}
\end{table}

\addtocounter{table}{-1}

\begin{table}[htb]
\begin{center}
\caption{-- \it continued}
\bigskip
\footnotesize
\begin{tabular}{cc|cc|cc|cc}
\tableline\tableline
\multicolumn{2}{c}{1.4 GHz} & \multicolumn{2}{c}{2.4 GHz} & 
\multicolumn{2}{c}{4.8 GHz} & \multicolumn{2}{c}{8.6 GHz} \\

\tableline
Day & Flux Density & Day & Flux Density & Day & Flux Density & Day & Flux Density \\
 	& (mJy) 	       &  	 & (mJy) 		&   	   & (mJy) 		  &  	    & (mJy) \\ 
\tableline

-	&		-		&	-	&		-		&	1787	& $	7.4	\pm	0.4	$ &	1787	& $	3.9	\pm	0.3	$ \\
1789	& $	20.6	\pm	1.4	$ &	1789	& $	10.7	\pm	1.1	$ &	-	&		-		&	-	&		-		\\
1850	& $	24.8	\pm	1.4	$ &	1850	& $	13.1	\pm	1.1	$ &	-	&		-		&	-	&		-		\\
-	&		-		&	-	&		-		&	1852	& $	8.2	\pm	0.4	$ &	1852	& $	4.4	\pm	0.3	$ \\
1879	& $	24.0	\pm	1.4	$ &	1879	& $	15.0	\pm	1.1	$ &	1878	& $	7.1	\pm	0.4	$ &	1878	& $	4.1	\pm	0.3	$ \\
-	&		-		&	-	&		-		&	1948	& $	8.3	\pm	0.4	$ &	1948	& $	4.7	\pm	0.3	$ \\
1970	& $	27.5	\pm	1.5	$ &	1970	& $	18.0	\pm	1.1	$ &	1970	& $	9.1	\pm	0.4	$ &	1970	& $	4.9	\pm	0.4	$ \\
-	&		-		&	-	&		-		&	1986	& $	8.4	\pm	0.4	$ &	1986	& $	4.6	\pm	0.3	$ \\
-	&		-		&	-	&		-		&	2003	& $	8.7	\pm	0.4	$ &	2003	& $	5.1	\pm	0.4	$ \\
2068	& $	33.2	\pm	1.6	$ &	2068	& $	19.0	\pm	1.2	$ &	2067	& $	9.4	\pm	0.5	$ &	2068	& $	5.5	\pm	0.4	$ \\
-	&		-		&	-	&		-		&	2092	& $	9.9	\pm	0.5	$ &	2092	& $	5.9	\pm	0.4	$ \\
-	&		-		&	-	&		-		&	2143	& $	8.6	\pm	0.4	$ &	2142	& $	5.4	\pm	0.4	$ \\
-	&		-		&	-	&		-		&	2196	& $	9.1	\pm	0.4	$ &	2196	& $	4.2	\pm	0.3	$ \\
-	&		-		&	-	&		-		&	2262	& $	10.9	\pm	0.5	$ &	2262	& $	6.0	\pm	0.4	$ \\
-	&		-		&	-	&		-		&	2300	& $	11.5	\pm	0.5	$ &	2300	& $	7.1	\pm	0.4	$ \\
-	&		-		&	-	&		-		&	2300	& $	11.6	\pm	0.5	$ &	2300	& $	5.8	\pm	0.4	$ \\
2314	& $	38.9	\pm	1.7	$ &	2314	& $	23.3	\pm	1.2	$ &	-	&		-		&	2314	& $	6.9	\pm	0.4	$ \\
-	&		-		&	-	&		-		&	-	&		-		&	2321	& $	6.9	\pm	0.4	$ \\
-	&		-		&	-	&		-		&	2375	& $	10.8	\pm	0.5	$ &	2375	& $	5.4	\pm	0.4	$ \\
-	&		-		&	-	&		-		&	2404	& $	12.4	\pm	0.6	$ &	2404	& $	7.1	\pm	0.4	$ \\
-	&		-		&	-	&		-		&	-	&		-		&	2426	& $	7.7	\pm	0.5	$ \\
-	&		-		&	-	&		-		&	2463	& $	12.8	\pm	0.6	$ &	2463	& $	6.5	\pm	0.4	$ \\
2505	& $	43.2	\pm	1.8	$ &	2505	& $	25.9	\pm	1.3	$ &	-	&		-		&	-	&		-		\\
-	&		-		&	-	&		-		&	2511	& $	12.0	\pm	0.5	$ &	-	&		-		\\
2549	& $	43.8	\pm	1.8	$ &	2549	& $	26.6	\pm	1.3	$ &	-	&		-		&	2550	& $	7.5	\pm	0.5	$ \\
-	&		-		&	-	&		-		&	2572	& $	14.0	\pm	0.6	$ &	2572	& $	5.9	\pm	0.4	$ \\
-	&		-		&	-	&		-		&	2580	& $	13.6	\pm	0.6	$ &	2580	& $	6.9	\pm	0.4	$ \\
-	&		-		&	-	&		-		&	2580	& $	13.1	\pm	0.6	$ &	2580	& $	6.5	\pm	0.4	$ \\
-	&		-		&	-	&		-		&	2628	& $	13.5	\pm	0.6	$ &	2628	& $	9.0	\pm	0.5	$ \\
-	&		-		&	-	&		-		&	2628	& $	14.1	\pm	0.6	$ &	2628	& $	7.8	\pm	0.5	$ \\
-	&		-		&	-	&		-		&	2648	& $	12.9	\pm	0.6	$ &	2648	& $	6.1	\pm	0.4	$ \\
2681	& $	46.2	\pm	1.8	$ &	2681	& $	28.4	\pm	1.3	$ &	-	&		-		&	-	&		-		\\
-	&		-		&	-	&		-		&	2754	& $	13.6	\pm	0.6	$ &	2754	& $	5.4	\pm	0.4	$ \\
-	&		-		&	-	&		-		&	2755	& $	14.2	\pm	0.6	$ &	2755	& $	6.3	\pm	0.4	$ \\
2774	& $	48.6	\pm	1.9	$ &	2774	& $	30.4	\pm	1.4	$ &	2774	& $	14.1	\pm	0.6	$ &	-	&		-		\\
2775	& $	49.3	\pm	1.9	$ &	2775	& $	29.9	\pm	1.3	$ &	2775	& $	14.6	\pm	0.6	$ &	2775	& $	9.2	\pm	0.5	$ \\

\tableline
\end{tabular}
\end{center}
\end{table}

\addtocounter{table}{-1}

\begin{table}[htb]
\begin{center}
\caption{-- \it continued}
\bigskip
\footnotesize
\begin{tabular}{cc|cc|cc|cc}
\tableline\tableline
\multicolumn{2}{c}{1.4 GHz} & \multicolumn{2}{c}{2.4 GHz} & 
\multicolumn{2}{c}{4.8 GHz} & \multicolumn{2}{c}{8.6 GHz} \\

\tableline
Day & Flux Density & Day & Flux Density & Day & Flux Density & Day & Flux Density \\
 	& (mJy) 	       &  	 & (mJy) 		&   	   & (mJy) 		  &  	    & (mJy) \\ 
\tableline

-	&		-		&	2827	& $	28.4	\pm	1.3	$ &	2827	& $	14.5	\pm	0.6	$ &	2827	& $	9.8	\pm	0.6	$ \\
2858	& $	50.9	\pm	1.9	$ &	2858	& $	29.3	\pm	1.3	$ &	-	&		-		&	-	&		-		\\
2874	& $	48.4	\pm	1.9	$ &	2874	& $	25.2	\pm	1.3	$ &	-	&		-		&	-	&		-		\\
2919	& $	53.8	\pm	2.0	$ &	2919	& $	33.1	\pm	1.4	$ &	2919	& $	16.5	\pm	0.7	$ &	2919	& $	9.4	\pm	0.5	$ \\
2919	& $	51.4	\pm	2.0	$ &	2919	& $	34.2	\pm	1.4	$ &	2919	& $	16.1	\pm	0.7	$ &	2919	& $	11.3	\pm	0.6	$ \\
2976	& $	55.0	\pm	2.0	$ &	2976	& $	33.6	\pm	1.4	$ &	2976	& $	15.3	\pm	0.7	$ &	2976	& $	9.1	\pm	0.5	$ \\
2976	& $	48.4	\pm	1.9	$ &	2976	& $	32.5	\pm	1.4	$ &	2976	& $	16.5	\pm	0.7	$ &	2976	& $	8.3	\pm	0.5	$ \\
3001	& $	56.2	\pm	2.1	$ &	3002	& $	33.1	\pm	1.4	$ &	3001	& $	15.1	\pm	0.7	$ &	3001	& $	6.6	\pm	0.4	$ \\
3002	& $	48.8	\pm	1.9	$ &	3073	& $	34.6	\pm	1.4	$ &	3002	& $	16.8	\pm	0.7	$ &	3002	& $	9.2	\pm	0.5	$ \\
3073	& $	57.2	\pm	2.1	$ &	3140	& $	36.6	\pm	1.5	$ &	3073	& $	17.8	\pm	0.8	$ &	3073	& $	10.4	\pm	0.6	$ \\
-	&		-		&	-	&		-		&	-	&		-		&	3074	& $	11.4	\pm	0.6	$ \\
-	&		-		&	-	&		-		&	-	&		-		&	3111	& $	10.2	\pm	0.6	$ \\
3177	& $	59.1	\pm	2.1	$ &	3177	& $	35.9	\pm	1.5	$ &	3177	& $	16.9	\pm	0.7	$ &	3177	& $	9.8	\pm	0.6	$ \\
3203	& $	64.2	\pm	2.3	$ &	3203	& $	39.2	\pm	1.5	$ &	3203	& $	20.2	\pm	0.9	$ &	3203	& $	12.6	\pm	0.7	$ \\
3278	& $	66.2	\pm	2.3	$ &	3278	& $	41.6	\pm	1.6	$ &	3278	& $	21.6	\pm	0.9	$ &	3278	& $	12.1	\pm	0.7	$ \\
3326	& $	67.1	\pm	2.3	$ &	3326	& $	41.4	\pm	1.6	$ &	3326	& $	20.7	\pm	0.9	$ &	3326	& $	11.7	\pm	0.6	$ \\
3415	& $	71.8	\pm	2.5	$ &	3415	& $	44.1	\pm	1.7	$ &	3415	& $	21.3	\pm	0.9	$ &	3415	& $	11.5	\pm	0.6	$ \\
-	&		-		&	-	&		-		&	-	&		-		&	3437	& $	15.7	\pm	0.8	$ \\
3456	& $	71.8	\pm	2.5	$ &	3456	& $	44.8	\pm	1.7	$ &	3456	& $	23.1	\pm	1.0	$ &	3456	& $	14.2	\pm	0.8	$ \\
-	&		-		&	-	&		-		&	-	&		-		&	3486	& $	16.1	\pm	0.8	$ \\
-	&		-		&	-	&		-		&	-	&		-		&	3512	& $	16.5	\pm	0.9	$ \\
3515	& $	73.3	\pm	2.5	$ &	3515	& $	45.8	\pm	1.7	$ &	3515	& $	24.8	\pm	1.0	$ &	3515	& $	14.6	\pm	0.8	$ \\
3579	& $	77.5	\pm	2.6	$ &	3579	& $	48.5	\pm	1.8	$ &	3579	& $	25.4	\pm	1.1	$ &	3579	& $	15.5	\pm	0.8	$ \\
3633	& $	79.1	\pm	2.7	$ &	3633	& $	48.0	\pm	1.8	$ &	3633	& $	25.1	\pm	1.0	$ &	3633	& $	14.9	\pm	0.8	$ \\
3679	& $	79.6	\pm	2.7	$ &	3679	& $	49.4	\pm	1.8	$ &	3679	& $	24.5	\pm	1.0	$ &	3679	& $	16.3	\pm	0.9	$ \\
3714	& $	79.3	\pm	2.7	$ &	3714	& $	50.1	\pm	1.8	$ &	3714	& $	24.5	\pm	1.0	$ &	3714	& $	15.9	\pm	0.8	$ \\
3745	& $	84.4	\pm	2.8	$ &	3745	& $	48.5	\pm	1.8	$ &	3745	& $	27.0	\pm	1.1	$ &	3745	& $	16.3	\pm	0.9	$ \\
3772	& $	81.9	\pm	2.7	$ &	3772	& $	49.9	\pm	1.8	$ &	3772	& $	27.1	\pm	1.1	$ &	3772	& $	17.9	\pm	0.9	$ \\
3834	& $	84.2	\pm	2.8	$ &	3834	& $	50.1	\pm	1.8	$ &	3834	& $	26.5	\pm	1.1	$ &	3834	& $	17.7	\pm	0.9	$ \\
3900	& $	86.0	\pm	2.9	$ &	3900	& $	52.6	\pm	1.9	$ &	3900	& $	29.2	\pm	1.2	$ &	3900	& $	17.1	\pm	0.9	$ \\
3945	& $	88.1	\pm	2.9	$ &	3945	& $	52.2	\pm	1.9	$ &	3945	& $	29.8	\pm	1.2	$ &	3945	& $	17.9	\pm	0.9	$ \\
3987	& $	93.8	\pm	3.1	$ &	3987	& $	52.4	\pm	1.9	$ &	3987	& $	31.9	\pm	1.3	$ &	3987	& $	18.9	\pm	1.0	$ \\
4015	& $	89.3	\pm	2.9	$ &	4015	& $	55.3	\pm	1.9	$ &	4015	& $	30.2	\pm	1.2	$ &	4015	& $	19.4	\pm	1.0	$ \\
4059	& $	87.7	\pm	2.9	$ &	4059	& $	58.3	\pm	2.0	$ &	-	&		-		&	-	&		-		\\
4101	& $	96.7	\pm	3.1	$ &	4101	& $	58.3	\pm	2.0	$ &	4101	& $	32.0	\pm	1.3	$ &	4101	& $	19.6	\pm	1.0	$ \\
4170	& $	99.6	\pm	3.2	$ &	4170	& $	61.4	\pm	2.1	$ &	-	&		-		&	-	&		-		\\

\tableline
\end{tabular}
\end{center}
\end{table}

\addtocounter{table}{-1}

\begin{table}[htb]
\begin{center}
\caption{-- \it continued}
\bigskip
\footnotesize
\begin{tabular}{cc|cc|cc|cc}
\tableline\tableline
\multicolumn{2}{c}{1.4 GHz} & \multicolumn{2}{c}{2.4 GHz} & 
\multicolumn{2}{c}{4.8 GHz} & \multicolumn{2}{c}{8.6 GHz} \\

\tableline
Day & Flux Density & Day & Flux Density & Day & Flux Density & Day & Flux Density \\
 	& (mJy) 	       &  	 & (mJy) 		&   	   & (mJy) 		  &  	    & (mJy) \\ 
\tableline

4223	& $	97.5	\pm	3.2	$ &	4223	& $	59.2	\pm	2.0	$ &	4223	& $	32.0	\pm	1.3	$ &	4223	& $	19.1	\pm	1.0	$ \\
4291	& $	104.5	\pm	3.4	$ &	4291	& $	64.7	\pm	2.2	$ &	4291	& $	35.3	\pm	1.4	$ &	4291	& $	21.0	\pm	1.1	$ \\
4373	& $	108.3	\pm	3.5	$ &	4373	& $	62.3	\pm	2.1	$ &	-	&		-		&	4373	& $	21.2	\pm	1.1	$ \\
4424	& $	107.4	\pm	3.4	$ &	4424	& $	67.1	\pm	2.3	$ &	4424	& $	34.1	\pm	1.4	$ &	4424	& $	24.9	\pm	1.3	$ \\
4461	& $	113.2	\pm	3.6	$ &	4461	& $	68.9	\pm	2.3	$ &	4461	& $	38.6	\pm	1.6	$ &	4461	& $	25.5	\pm	1.3	$ \\
4540	& $	112.6	\pm	3.6	$ &	4540	& $	75.5	\pm	2.5	$ &	4540	& $	37.6	\pm	1.5	$ &	4540	& $	25.0	\pm	1.3	$ \\
4572	& $	116.1	\pm	3.7	$ &	4572	& $	69.5	\pm	2.3	$ &	4572	& $	39.6	\pm	1.6	$ &	4572	& $	24.4	\pm	1.3	$ \\
4685	& $	121.2	\pm	3.8	$ &	4685	& $	75.2	\pm	2.5	$ &	4685	& $	40.1	\pm	1.6	$ &	4685	& $	26.7	\pm	1.4	$ \\
4729	& $	121.4	\pm	3.8	$ &	4729	& $	74.2	\pm	2.4	$ &	4729	& $	43.7	\pm	1.8	$ &	4729	& $	28.1	\pm	1.4	$ \\
4768	& $	131.3	\pm	4.1	$ &	4768	& $	77.1	\pm	2.5	$ &	4768	& $	42.7	\pm	1.7	$ &	4768	& $	25.7	\pm	1.3	$ \\
4800	& $	124.9	\pm	3.9	$ &	4800	& $	77.4	\pm	2.5	$ &	4800	& $	39.7	\pm	1.6	$ &	4800	& $	24.8	\pm	1.3	$ \\
4838	& $	126.4	\pm	4.0	$ &	4838	& $	83.8	\pm	2.7	$ &	4838	& $	43.6	\pm	1.8	$ &	4838	& $	31.5	\pm	1.6	$ \\
4851	& $	128.7	\pm	4.0	$ &	4851	& $	77.8	\pm	2.5	$ &	4851	& $	45.5	\pm	1.8	$ &	4851	& $	27.9	\pm	1.4	$ \\
4871	& $	127.0	\pm	4.0	$ &	4871	& $	77.9	\pm	2.5	$ &	4871	& $	47.2	\pm	1.9	$ &	4871	& $	29.9	\pm	1.5	$ \\
4937	& $	133.4	\pm	4.2	$ &	4937	& $	81.0	\pm	2.6	$ &	4937	& $	48.1	\pm	1.9	$ &	4937	& $	30.2	\pm	1.5	$ \\
4997	& $	136.9	\pm	4.3	$ &	4997	& $	83.6	\pm	2.7	$ &	4997	& $	43.5	\pm	1.8	$ &	4997	& $	27.5	\pm	1.4	$ \\
5025	& $	135.0	\pm	4.2	$ &	5025	& $	83.0	\pm	2.7	$ &	5025	& $	47.7	\pm	1.9	$ &	5025	& $	31.9	\pm	1.6	$ \\
5050	& $	135.9	\pm	4.3	$ &	5050	& $	86.2	\pm	2.8	$ &	5050	& $	46.4	\pm	1.9	$ &	5050	& $	31.9	\pm	1.6	$ \\
5093	& $	143.4	\pm	4.5	$ &	5093	& $	86.6	\pm	2.8	$ &	5093	& $	49.9	\pm	2.0	$ &	5093	& $	36.0	\pm	1.8	$ \\
5128	& $	139.6	\pm	4.4	$ &	5128	& $	83.5	\pm	2.7	$ &	5128	& $	48.8	\pm	2.0	$ &	5128	& $	30.3	\pm	1.5	$ \\
5160	& $	144.7	\pm	4.5	$ &	5160	& $	88.7	\pm	2.8	$ &	5160	& $	50.7	\pm	2.0	$ &	5160	& $	29.6	\pm	1.5	$ \\
5293	& $	153.3	\pm	4.8	$ &	5293	& $	95.9	\pm	3.1	$ &	5293	& $	53.1	\pm	2.1	$ &	5293	& $	31.7	\pm	1.6	$ \\
5369	& $	155.1	\pm	4.8	$ &	5369	& $	98.5	\pm	3.1	$ &	5369	& $	56.4	\pm	2.3	$ &	5369	& $	36.9	\pm	1.9	$ \\
5401	& $	154.9	\pm	4.8	$ &	5401	& $	96.0	\pm	3.1	$ &	5401	& $	54.6	\pm	2.2	$ &	5401	& $	34.2	\pm	1.7	$ \\
5460	& $	160.8	\pm	5.0	$ &	5460	& $	98.1	\pm	3.1	$ &	5524	& $	60.2	\pm	2.4	$ &	5524	& $	39.3	\pm	2.0	$ \\
5524	& $	166.9	\pm	5.2	$ &	5524	& $	103.0 \pm	3.3	$ &	5546	& $	62.6	\pm	2.5	$ &	5546	& $	41.4	\pm	2.1	$ \\
5546	& $	166.4	\pm	5.1	$ &	5546	& $	100.2 \pm	3.2	$ &	5574	& $	62.7	\pm	2.5	$ &	5574	& $	42.7	\pm	2.2	$ \\
5574	& $	175.5	\pm	5.4	$ &	5574	& $	103.8 \pm	3.3	$ &	5668	& $	57.1	\pm	2.3	$ &	5668	& $	48.8	\pm	2.5	$ \\
5729	& $	179.7	\pm	5.5	$ &	5729	& $	109.7 \pm	3.4	$ &	5729	& $	65.6	\pm	2.6	$ &	5729	& $	41.3	\pm	2.1	$ \\
-	&		-			    &	5774	& $	112.0 \pm	3.5	$ &	-	&		-		&	5774	& $	48.6	\pm	2.4	$ \\
5798	& $	176.3	\pm	5.4	$ &	5798	& $	114.8 \pm	3.6	$ &	5798	& $	62.3	\pm	2.5	$ &	5798	& $	41.4	\pm	2.1	$ \\
5851	& $	184.0	\pm	5.7	$ &	5851	& $	120.0 \pm	3.7	$ &	5851	& $	66.5	\pm	2.7	$ &	5851	& $	49.3	\pm	2.5	$ \\
5930	& $	190.5	\pm	5.8	$ &	5930	& $	124.7 \pm	3.9	$ &	5930	& $	68.1	\pm	2.7	$ &	5930	& $	44.2	\pm	2.2	$ \\
5970	& $	195.0	\pm	6.0	$ &	5970	& $	115.5 \pm	3.6	$ &	-	&		-		&	-	&		-		\\
5980	& $	197.0	\pm	6.0	$ &	-	&		-		&	-	&		-		&	-	&		-		\\
6005	& $	203.2	\pm	6.2	$ &	6005	& $	130.7 \pm	4.1	$ &	6005	& $	76.0	\pm	3.1	$ &	6005	& $	48.9	\pm	2.5	$ \\

\tableline
\end{tabular}
\end{center}
\end{table}

\addtocounter{table}{-1}

\begin{table}[htb]
\begin{center}
\caption{-- \it continued}
\bigskip
\footnotesize
\begin{tabular}{cc|cc|cc|cc}
\tableline\tableline
\multicolumn{2}{c}{1.4 GHz} & \multicolumn{2}{c}{2.4 GHz} & 
\multicolumn{2}{c}{4.8 GHz} & \multicolumn{2}{c}{8.6 GHz} \\

\tableline
Day & Flux Density & Day & Flux Density & Day & Flux Density & Day & Flux Density \\
 	& (mJy) 	       &  	 & (mJy) 		&   	   & (mJy) 		  &  	    & (mJy) \\ 
\tableline

6139	& $	206.5	\pm	6.3	$ &	6139	& $	133.9 \pm	4.1	$ &	6139	& $	82.4	\pm	3.3	$ &	6139	& $	54.0	\pm	2.7	$ \\
6172	& $	208.7	\pm	6.4	$ &	6172	& $	132.9 \pm	4.1	$ &	-	&		-		&	-	&		-		\\
6199	& $	208.5	\pm	6.4	$ &	6199	& $	136.2 \pm	4.2	$ &	6199	& $	73.4	\pm	3.0	$ &	6199	& $	49.6	\pm	2.5	$ \\
6244	& $	217.4	\pm	6.6	$ &	6244	& $	139.1 \pm	4.3	$ &	6244	& $	76.8	\pm	3.1	$ &	6244	& $	54.9	\pm	2.8	$ \\
6283	& $	213.0	\pm	6.5	$ &	6283	& $	138.5	\pm	4.3	$ &	6283	& $	83.7	\pm	3.4	$ &	6283	& $	52.4	\pm	2.6	$ \\
6355	& $	224.9	\pm	6.9	$ &	6355	& $	148.6	\pm	4.6	$ &	6355	& $	86.8	\pm	3.5	$ &	6355	& $	54.3	\pm	2.7	$ \\
6461	& $	235.0	\pm	7.2	$ &	6461	& $	154.4	\pm	4.7	$ &	6461	& $	78.9	\pm	3.2	$ &	6461	& $	52.9	\pm	2.7	$ \\
6494	& $	242.8	\pm	7.4	$ &	6494	& $	160.0	\pm	4.9	$ &	6494	& $	96.3	\pm	3.9	$ &	6494	& $	56.4	\pm	2.8	$ \\
6526	& $	237.0	\pm	7.2	$ &	6526	& $	160.4	\pm	4.9	$ &	6526	& $	89.8	\pm	3.6	$ &	6526	& $	65.8	\pm	3.3	$ \\
6603	& $	252.6	\pm	7.7	$ &	6603	& $	165.4	\pm	5.1	$ &	6603	& $	102.0	\pm	4.1	$ &	6603	& $	62.6	\pm	3.1	$ \\
6627	& $	251.5	\pm	7.6	$ &	6627	& $	165.7	\pm	5.1	$ &	6627	& $	91.5	\pm	3.7	$ &	6627	& $	63.9	\pm	3.2	$ \\
6690	& $	251.5	\pm	7.6	$ &	6690	& $	171.0	\pm	5.2	$ &	6690	& $	100.9	\pm	4.0	$ &	6690	& $	62.5	\pm	3.1	$ \\
6770	& $	258.6	\pm	7.9	$ &	6770	& $	175.7	\pm	5.4	$ &	6770	& $	104.6	\pm	4.2	$ &	6770	& $	66.9	\pm	3.4	$ \\
6836	& $	264.5	\pm	8.0	$ &	6836	& $	178.1	\pm	5.4	$ &	6836	& $	107.8	\pm	4.3	$ &	6836	& $	68.0	\pm	3.4	$ \\
6889	& $	272.5	\pm	8.3	$ &	6889	& $	184.8	\pm	5.6	$ &	6889	& $	111.7	\pm	4.5	$ &	6889	& $	72.1	\pm	3.6	$ \\
6967	& $	276.4	\pm	8.4	$ &	6967	& $	190.3	\pm	5.8	$ &	6967	& $	121.0	\pm	4.9	$ &	6967	& $	79.9	\pm	4.0	$ \\
6989	& $	281.3	\pm	8.5	$ &	6989	& $	179.7	\pm	5.5	$ &	6989	& $	118.1	\pm	4.7	$ &	6989	& $	81.8	\pm	4.1	$ \\
7039	& $	294.9	\pm	8.9	$ &	7039	& $	196.0	\pm	6.0	$ &	7039	& $	118.8	\pm	4.8	$ &	7039	& $	78.1	\pm	3.9	$ \\
7084	& $	287.1	\pm	8.7	$ &	7084	& $	194.7	\pm	5.9	$ &	7084	& $	123.1	\pm	4.9	$ &	7084	& $	78.3	\pm	3.9	$ \\
7202	& $	312.8	\pm	9.5	$ &	7202	& $	216.8	\pm	6.6	$ &	7202	& $	124.1	\pm	5.0	$ &	7202	& $	85.6	\pm	4.3	$ \\
7231	& $	304.4	\pm	9.2	$ &	7231	& $	206.2	\pm	6.3	$ &	7231	& $	130.4	\pm	5.2	$ &	7231	& $	83.9	\pm	4.2	$ \\
7283	& $	285.2	\pm	8.6	$ &	7283	& $	198.9	\pm	6.1	$ &	7283	& $	122.5	\pm	4.9	$ &	-	&		-		\\
7297	& $	315.2	\pm	9.5	$ &	7297	& $	213.0	\pm	6.5	$ &	7297	& $	136.8	\pm	5.5	$ &	7297	& $	90.1	\pm	4.5	$ \\
7370	& $	332.4	\pm	10.0	$ &	7370	& $	218.9	\pm	6.6	$ &	7371	& $	126.6	\pm	5.1	$ &	7371	& $	86.4	\pm	4.3	$ \\
7437	& $	342.5	\pm	10.3	$ &	7437	& $	229.5	\pm	7.0	$ &	7437	& $	140.4	\pm	5.6	$ &	7437	& $	92.3	\pm	4.6	$ \\
7486	& $	339.7	\pm	10.3	$ &	7486	& $	227.3	\pm	6.9	$ &	7486	& $	146.6	\pm	5.9	$ &	7486	& $	94.4	\pm	4.7	$ \\
7558	& $	355.8	\pm	10.7	$ &	7558	& $	244.3	\pm	7.4	$ &	7558	& $	137.9	\pm	5.5	$ &	7558	& $	105.3	\pm	5.3	$ \\
7580	& $	348.3	\pm	10.5	$ &	7580	& $	237.8	\pm	7.2	$ &	7580	& $	147.9	\pm	5.9	$ &	7580	& $	93.7	\pm	4.7	$ \\
7622	& $	361.0	\pm	10.9	$ &	7622	& $	239.5	\pm	7.3	$ &	7622	& $	154.3	\pm	6.2	$ &	7622	& $	97.7	\pm	4.9	$ \\
7692	& $	371.0	\pm	10.5	$ &	7692	& $	216.8	\pm	7.4	$ &	7692	& $	159.4	\pm	6.4	$ &	7692	& $	98.6	\pm	4.9	$ \\
7730	& $	379.2	\pm	11.4	$ &	7730	& $	259.5	\pm	7.9	$ &	7730	& $	165.5	\pm	6.6	$ &	7730	& $	107.5	\pm	5.4	$ \\
7802	& $	385.8	\pm	11.6	$ &	7802	& $	269.1	\pm	8.1	$ &	7802	& $	157.6	\pm	6.3	$ &	7802	& $	106.7	\pm	5.3	$ \\
7847	& $	389.9	\pm	11.8	$ &	7847	& $	268.9	\pm	8.1	$ &	7847	& $	167.9	\pm	6.7	$ &	7847	& $	106.9	\pm	5.4	$ \\
7901	& $	414.3	\pm	12.5	$ &	7901	& $	279.7	\pm	8.5	$ &	7900	& $	173.6	\pm	7.0	$ &	7900	& $	111.8	\pm	5.6	$ \\
7932	& $	414.6	\pm	12.5	$ &	7932	& $	283.8	\pm	8.6	$ &	7932	& $	177.9	\pm	7.1	$ &	7932	& $	113.0	\pm	5.7	$ \\
7991	& $	421.6	\pm	12.7	$ &	7991	& $	283.6	\pm	8.6	$ &	7991	& $	178.0	\pm	7.1	$ &	7991	& $	116.8	\pm	5.9	$ \\
8014	& $	421.5	\pm	12.7	$ &	8014	& $	298.9	\pm	9.0	$ &	8014	& $	178.2	\pm	7.1	$ &	8014	& $	119.6	\pm	6.0	$ \\

\tableline
\end{tabular}
\end{center}
\end{table}
\normalsize

\clearpage

\begin{table}[htb]
\begin{center}
\caption{Flux density exponential fitting parameters for ATCA and MOST\tablenotemark{1} data, from Day 5000.\label{ATCA fits}} 
\bigskip
\bigskip
\footnotesize
\begin{tabular}{crrc} 

\tableline\tableline

\multicolumn{1}{c}{$\nu$} & 
\multicolumn{1}{c}{$S_{0}$} & \multicolumn{1}{c}{$\Delta$} & \multicolumn{1}{c}{RMS}\\
\multicolumn{1}{c}{(GHz)}      & \multicolumn{1}{c}{(mJy)}     & \multicolumn{1}{c}{(day)}  &  \multicolumn{1}{c}{(mJy)}\\
\tableline

8.6--9			&	$30.8\pm0.7$	 &  $2231\pm63$		& 	3.18\\
4.8				&	$46.4\pm0.8$	 &  $2213\pm45$		& 	4.20\\
2.4				&	$82.8\pm1.0$	 &   $2417\pm40$ 	& 	6.44\\
1.4 				&	$135.3\pm1.1$	 &   $2678\pm34$ 	& 	6.23\\  
0.8				&	$207.6\pm1.7$	 &	$2502\pm42$  	& 	9.43\\

\tableline
\end{tabular}
\tablecomments{$S(t)=S_{0} e^{(t-t_{0})/\Delta}$, with $t$ expressed in days; $t_{0}=5000$ days; $\Delta$ is the $e$-folding time.}
\tablenotetext{1}{MOST flux densities at 843 MHz are available from Ball et al. (2001) and the web page
http://www.physics.usyd.edu.au/sifa/Main/SN1987A.}

\end{center}
\end{table}

\clearpage

\begin{table}[htb]
\begin{center}
\caption{Spectral index obtained by fitted flux densities from all observations between 843 MHz and 8.6 GHz ($\alpha_{1}$) and from ATCA-only observations between 1.4 and 8.6 GHz ($\alpha_{2}$). Day number is approximate.\label{MOST fits}} 
\bigskip

\bigskip
\footnotesize
\begin{tabular}{ccccrr}

\tableline\tableline

\multicolumn{1}{c}{Year} & 
\multicolumn{1}{c}{Day} & 
\multicolumn{1}{c}{$\alpha_{1}$} & 
\multicolumn{1}{c}{$\alpha_{2}$} & 
\multicolumn{1}{c}{$\Delta\alpha$\tablenotemark{1}} & 
\multicolumn{1}{c}{$\Delta\alpha/\alpha_{2}$} \\

\multicolumn{1}{c}{} & 
\multicolumn{1}{c}{} & 
\multicolumn{1}{c}{} & 
\multicolumn{1}{c}{} & 
\multicolumn{1}{c}{} & 
\multicolumn{1}{c}{(\%)} \\

\tableline

5	&	1517	& $	-0.932	\pm	0.051	$ & $	-0.885	\pm	0.068	$ & $	-0.047	 $ &	$5.26$\\
6	&	1878	& $	-0.994	\pm	0.021	$ & $	-0.975	\pm	0.028	$ & $	-0.019	 $ &	$1.96$\\
7	&	2300	& $	-1.012	\pm	0.025	$ & $	-1.033	\pm	0.034	$ & $	0.021	 $ &	$-2.03$\\
8	&	2774	& $	-0.923	\pm	0.020	$ & $	-0.927	\pm	0.033	$ & $	0.004	 $ &	$-0.46$\\
9	&	3203	& $	-0.892	\pm	0.014	$ & $	-0.895	\pm	0.023	$ & $	0.004	 $ &	$-0.41$\\
10	&	3633	& $	-0.928	\pm	0.011	$ & $	-0.913	\pm	0.010	$ & $	-0.015	 $ & $1.66$\\
11	&	4015	& $	-0.868	\pm	0.025	$ & $	-0.838	\pm	0.027	$ & $	-0.030	 $ &	$3.60$\\
12	&	4291	& $	-0.892	\pm	0.012	$ & $	-0.874	\pm	0.007	$ & $	-0.018	 $ &	$2.06$\\
13	&	4800	& $	-0.902	\pm	0.015	$ & $	-0.891	\pm	0.022	$ & $	-0.011	 $ &	$1.25$\\
14	&	5160	& $	-0.864	\pm	0.012	$ & $	-0.857	\pm	0.018	$ & $	-0.007	 $ &	$0.84$\\
15	&	5546	& $	-0.788	\pm	0.036	$ & $	-0.747	\pm	0.041	$ & $	-0.041	 $ &	$5.44$\\
16	&	6004	& $	-0.797	\pm	0.015	$ & $	-0.776	\pm	0.012	$ & $	-0.020	 $ &	$2.61$\\
17	&	6355	& $	-0.804	\pm	0.019	$ & $	-0.774	\pm	0.007	$ & $	-0.030	 $ &	$3.81$\\
18	&	6770	& $	-0.795	\pm	0.037	$ & $	-0.738	\pm	0.008	$ & $	-0.057	 $ &	$7.71$\\
19	&	7084	& $	-0.758	\pm	0.037	$ & $	-0.702	\pm	0.016	$ & $	-0.056	 $ &	$7.96$\\ 

\tableline
\end{tabular}
\tablenotetext{1}{$\Delta\alpha=\alpha_{2}-\alpha_{1}$. Note that the errors on $\alpha_{1}$ and $\alpha_{2}$ are correlated.}
\end{center}
\end{table}		


\begin{figure}[!ht]
\begin{center}
\centering
\includegraphics[width=8.8 cm, angle=270]{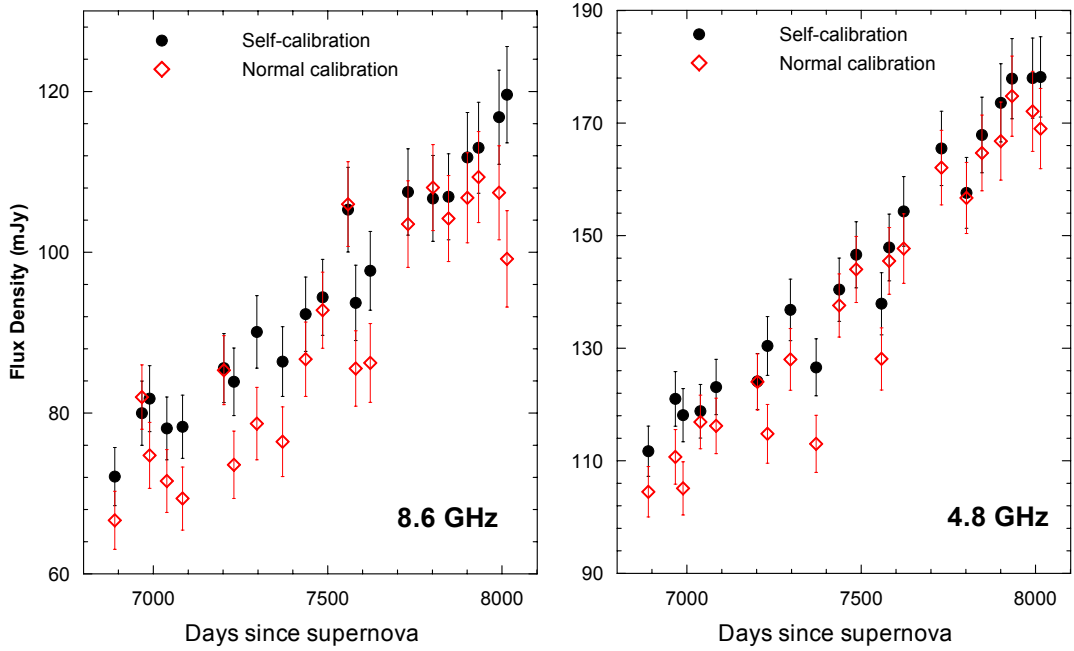}
\bigskip
\end{center}
\caption{Comparison of the flux densities derived from self-calibrated (solid black circles) and phase referenced  (open red diamonds) 4.8 and 8.6 GHz data, from days 6889 to 8014. Phase-only self-calibration increases the measured flux density and reduces the rms noise in images derived from observations made during times of poor phase stability (e.g. summer). 
\label{natural}}
\end{figure}

\clearpage

\begin{figure}[!ht]
\begin{center}
\includegraphics[width=7.5cm, angle=270]{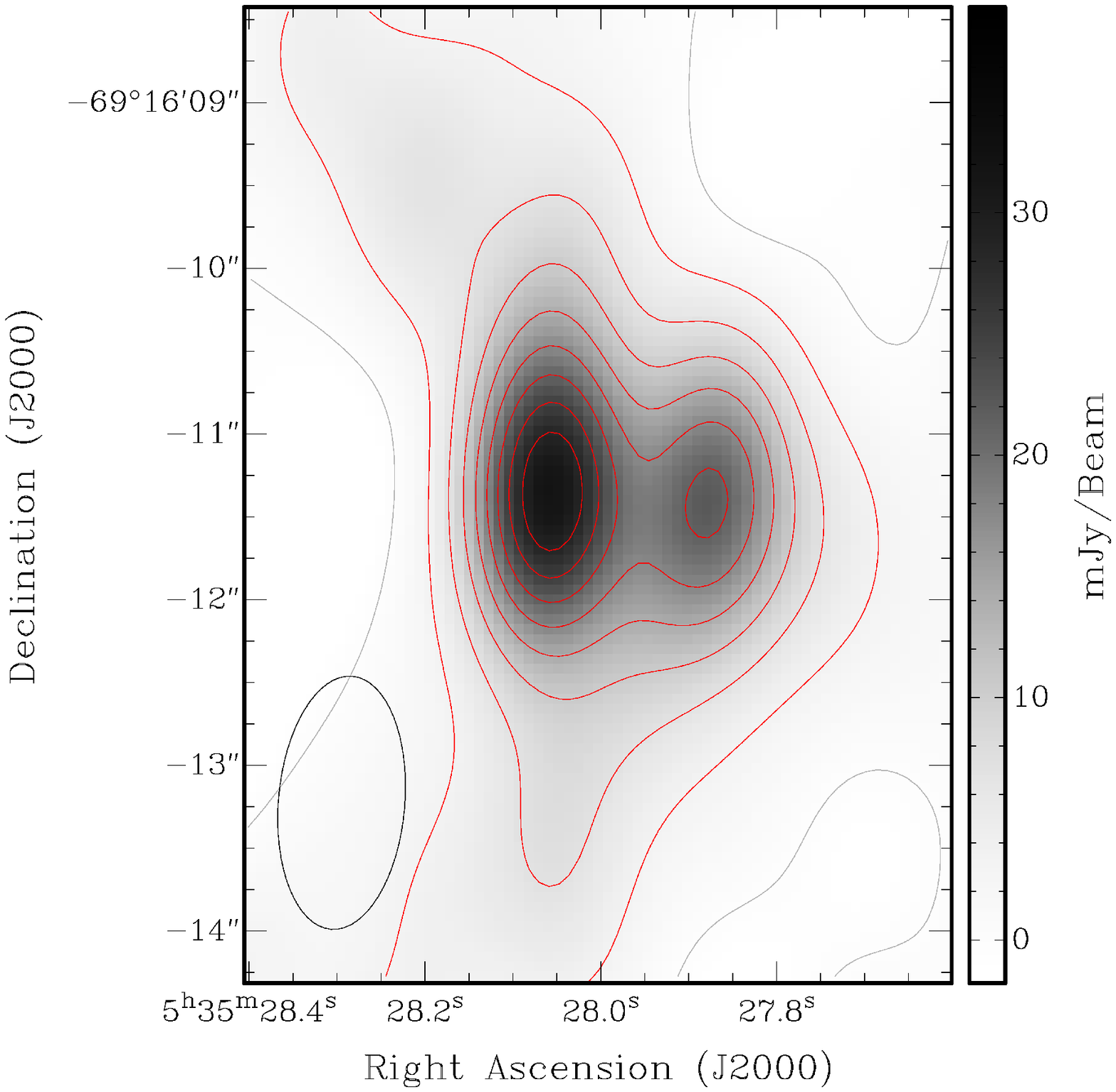}
\end{center}
\end{figure}

\begin{figure}[!ht]
\begin{center}
\includegraphics[width=7.5cm, angle=270]{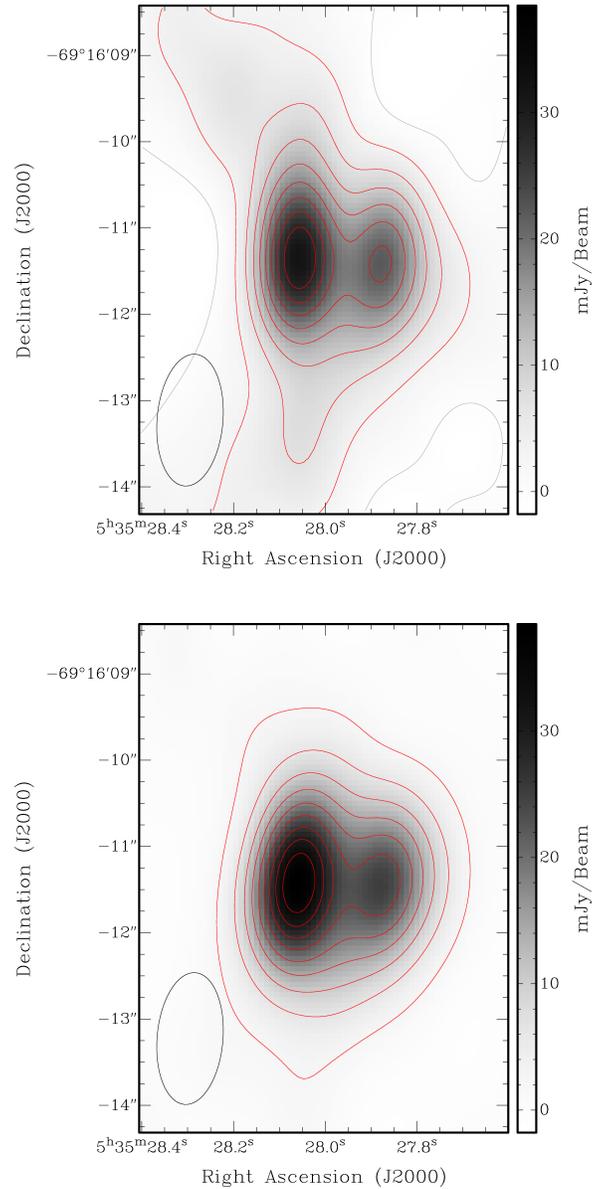}
\end{center}
\bigskip
\caption{Images of SN1987A at 8.6 GHz, as derived from monitoring observations taken on Day 7297 with the ATCA 6 km configuration (6A), before (top) and after (bottom) the application of the self-calibration procedure.  Monitoring typically involves 6--8 hrs of observations, rather than the full 12 hrs required for full $u-v$ coverage. While the normally calibrated image shows significant north-south artifacts and has a total flux of  78.7 mJy, the quality of the self-calibrated image is higher and the total flux is 90.1 mJy. The minimum contour level is at  0 mJy beam$^{-1}$ (gray line in the top figure), while the maximum is  at 34 mJy beam$^{-1}$  (most inner red line in the bottom figure), and the contour lines are spaced by  4 mJy beam$^{-1}$.
\label{natural}}
\end{figure}

\clearpage

\begin{figure}[!ht]
\centering
\includegraphics[width=19.5cm, angle=270]{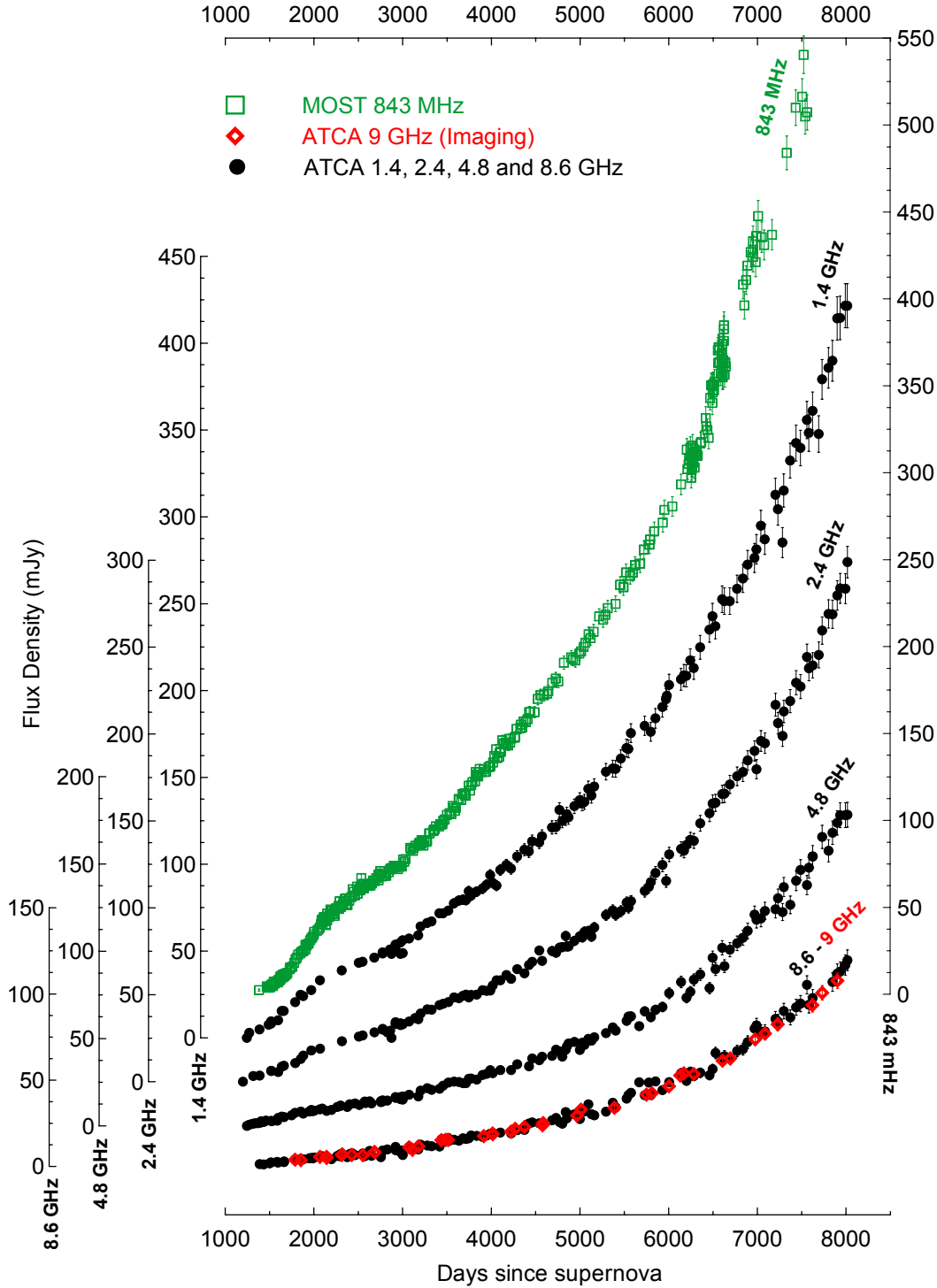}
\bigskip
\caption{Flux densities for SN 1987A, integrated over the whole remnant. Plots include data from observations at: 1.4, 2.4, 4.8 and 8.6 GHz from August 1990 to February 2009 (ATCA) (solid black circles); 9 GHz from January 1992 to October 2008 (ATCA) (open red diamonds); 843 MHz from February 1992 to March 2007 (MOST, Ball et al., 2001) (open green squares). The scale is linear and is the same for all frequencies apart from an offset of 25 mJy.
\label{natural}}
\end{figure}

\begin{figure}[!ht]
\centering
\includegraphics[width=18.5cm, angle=270]{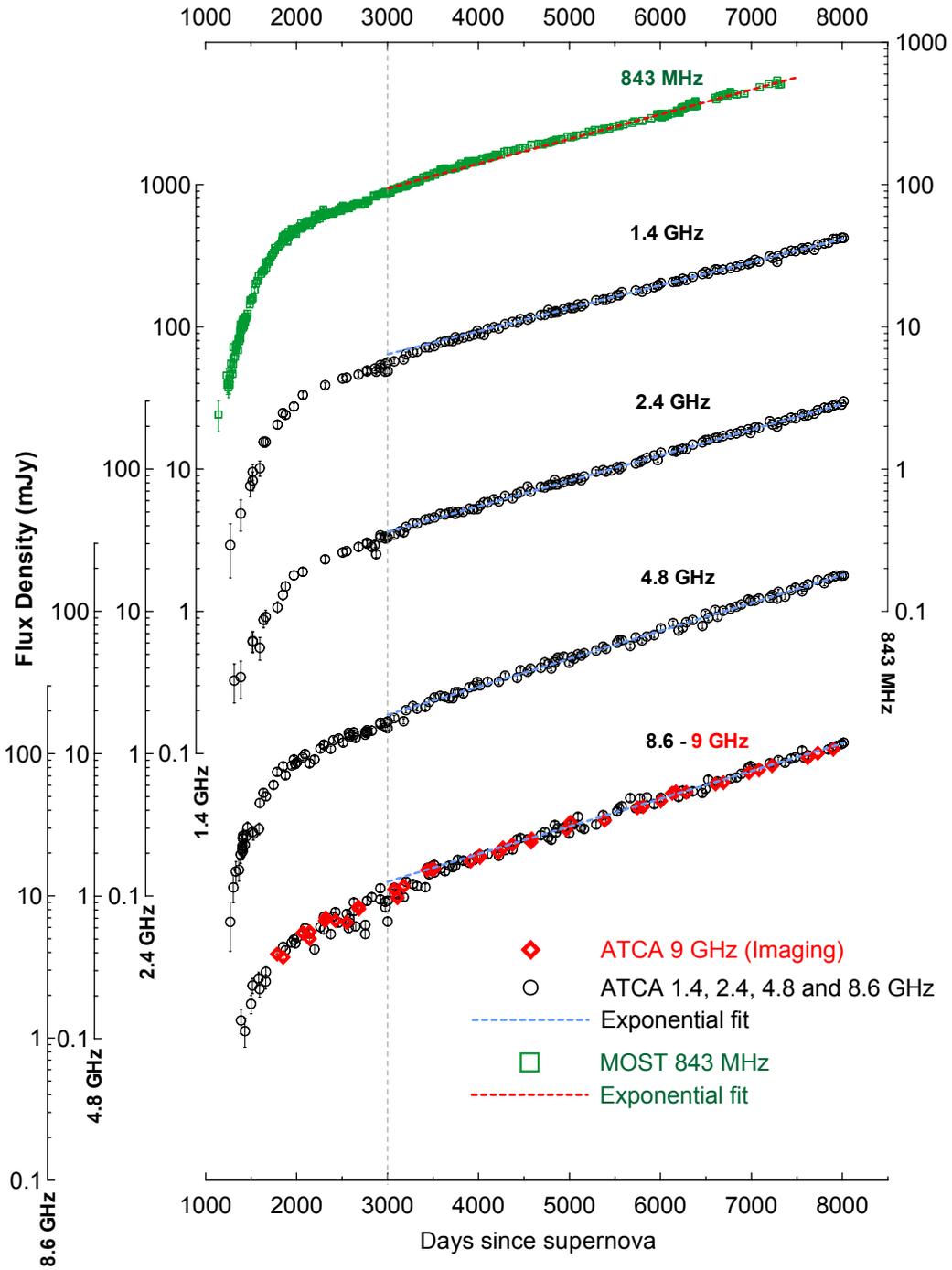}
\bigskip
\bigskip
\caption{Flux densities for SN 1987A. Plots include data from observations at: 1.4, 2.4, 4.8 and 8.6 GHz from August 1990 to February 2009 (ATCA) (solid black circles); 9 GHz from January 1992 to October 2008 (ATCA) (open red diamonds); 843 MHz from February 1992 to March 2007 (MOST, Ball et al., 2001) (open green squares). Fits parameters are given in Table 2. The scale is logarithmic and is the same for all frequencies apart from an offset of 0.1.
\label{natural}}
\end{figure}

\begin{figure}[!ht]
\centering
\includegraphics[width=11.5cm, angle=270]{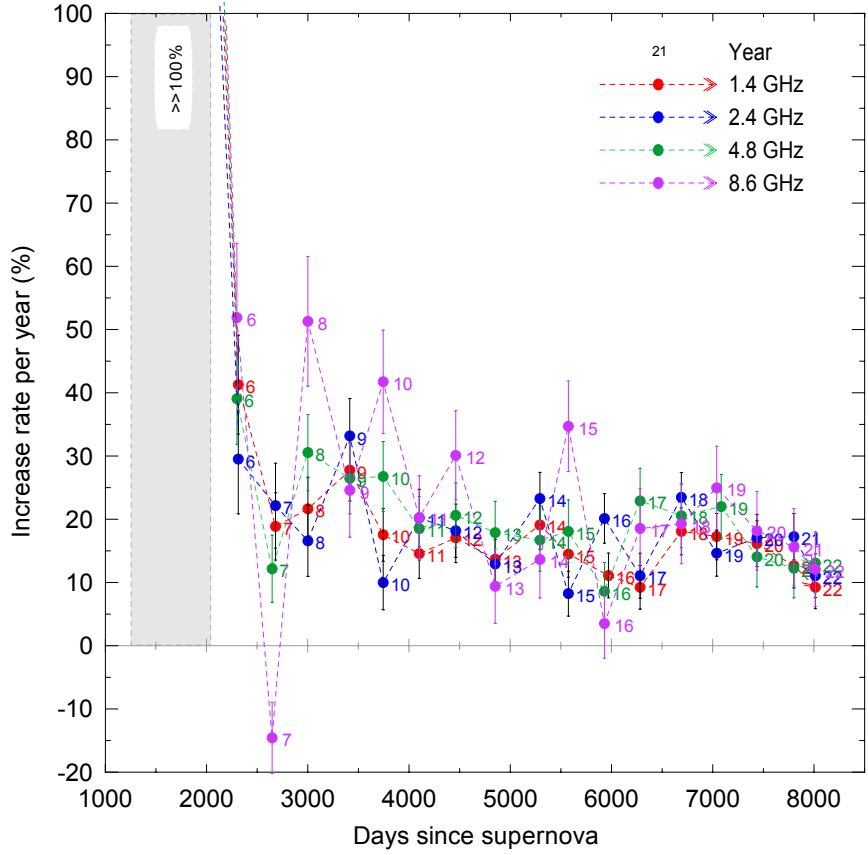}
\bigskip
\bigskip
\caption{Increase rate of flux densities per year, as derived from ATCA observations at 1.4, 2.4, 4.8 and 8.6 GHz from June 1990 to February 2009. Labels on nodal points indicate years since supernova. 
\label{natural}}
\end{figure}

\begin{figure}[!ht]
\centering
\includegraphics[width=15.7cm, angle=270]{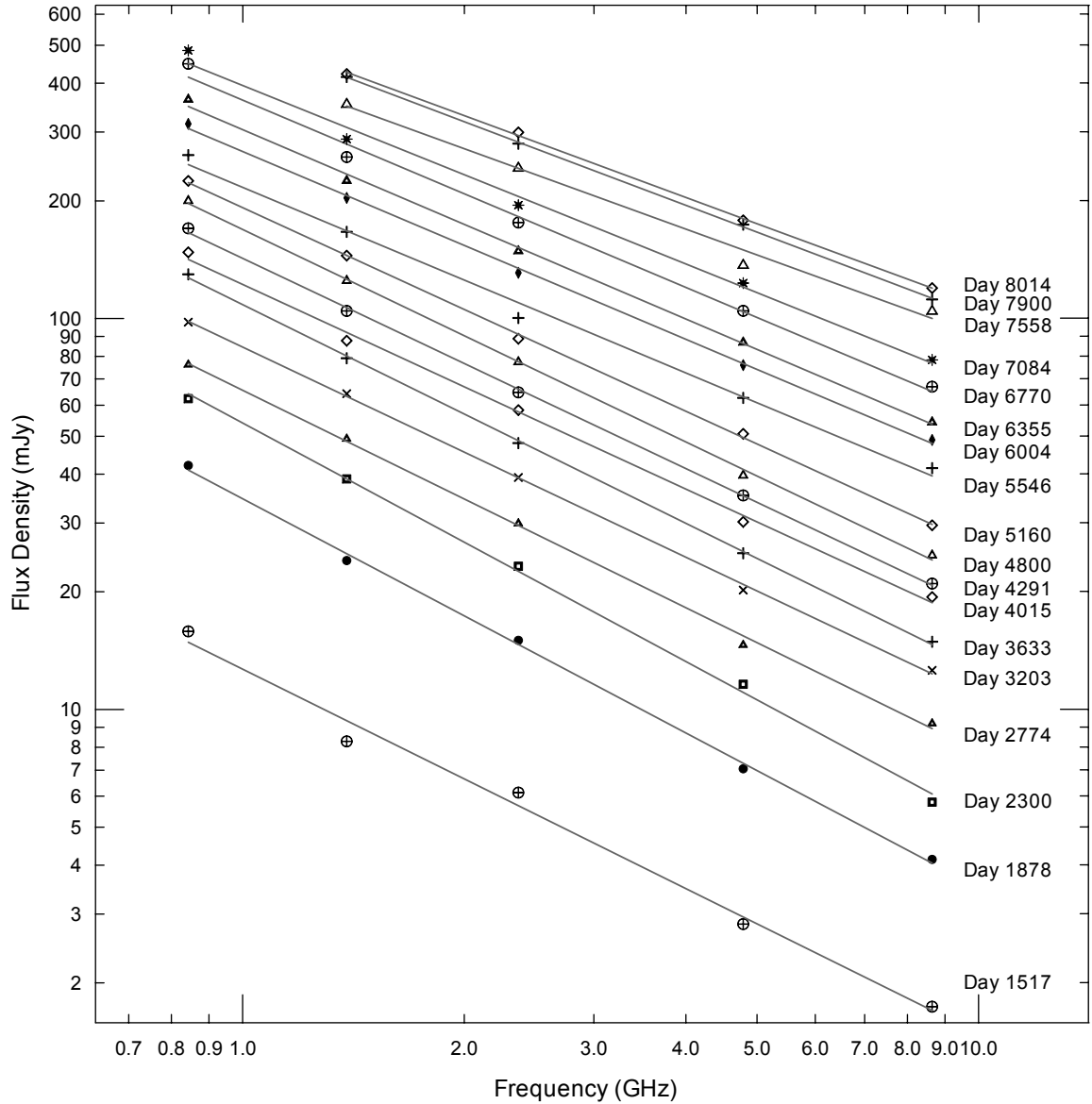}
\caption{Radio spectra from Year 4 to Year 22 since the supernova explosion at, approximately, yearly spaced epochs.
\label{natural}}
\end{figure}

\begin{figure}[!ht]
\centering
\includegraphics[width=10 cm, angle=270]{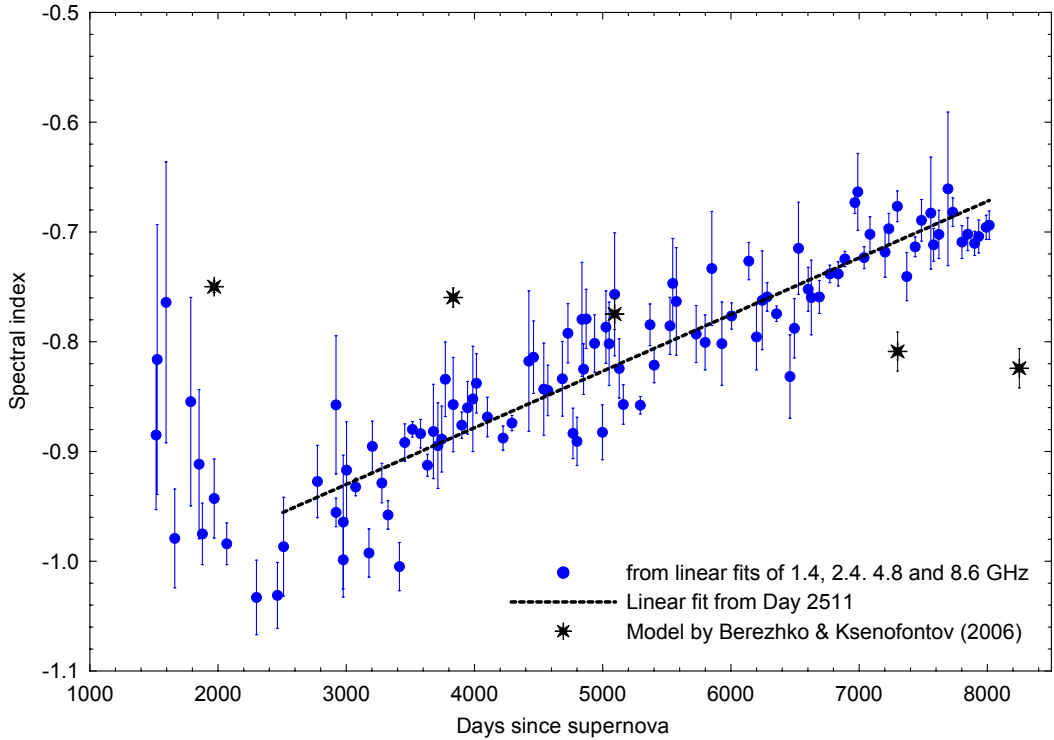}
\bigskip
\caption{Spectral indices as a function of time obtained from linear fits of ATCA observations at 1.4, 2.4, 4.8 and 8.6 GHz (full blue circles). The spectral index fit (dashed black line) is calculated from Day 2511 as $\alpha(t)=\alpha_{0} +\beta_{0}\times(t-t_{0})/\Delta$, where $t$ is expressed in days (d), $t_{0}=5000$ d and $\Delta=365$ d, $\alpha_{0}=-0.825\pm0.005$ and $\beta_{0}=0.018\pm0.001$. The black asterisks are from the model presented in  Berezhko \& Ksenofontov (2006), which corresponds to the upstream magnetic field $B_{1}\sim$ 3 mG and $\sigma\sim 3$. This model predicts a steepening of the spectral index over time instead of the observed flattening.
\label{natural}}
\end{figure}

\begin{figure}[!ht]
\centering
\includegraphics[width=10 cm, angle=270]{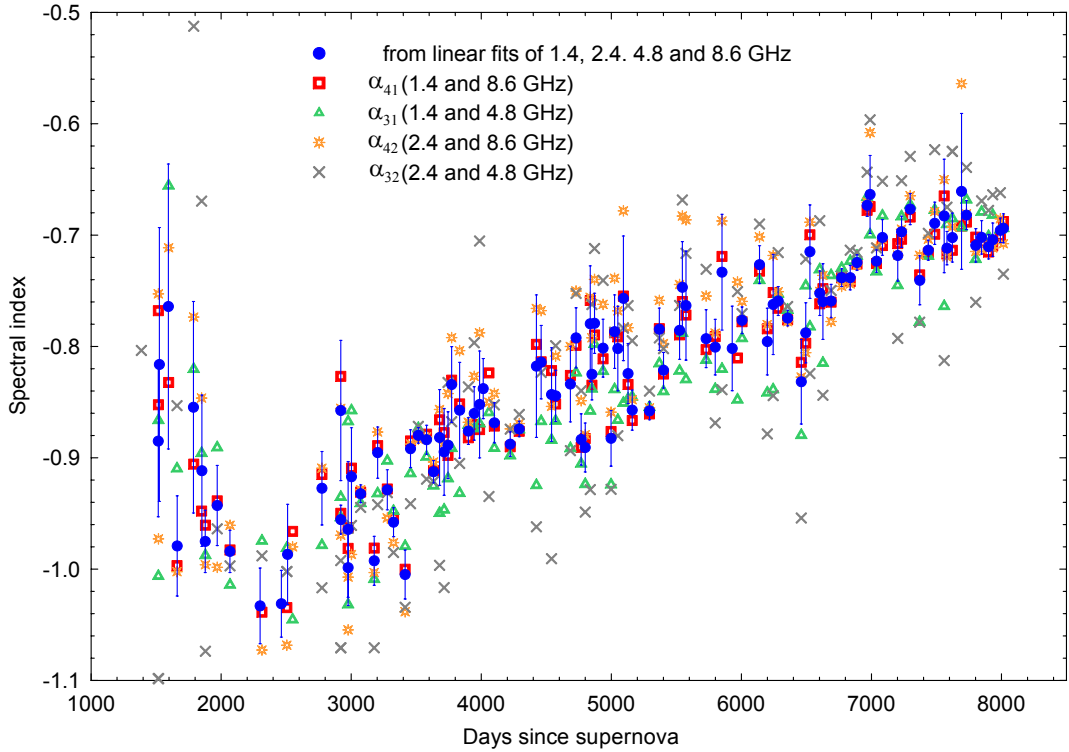}
\bigskip
\caption{Spectral indices derived from ratios between the higher (4.8 and 8.6 GHz) and lower (1.4 and 4.8 GHz) ATCA frequencies, are plotted as a function of time and compared to the values obtained from linear fits of radio spectra based on ATCA-only frequencies.
\label{natural}}
\end{figure}

\begin{figure}[!ht]
\centering
\includegraphics[width=11cm, angle=270]{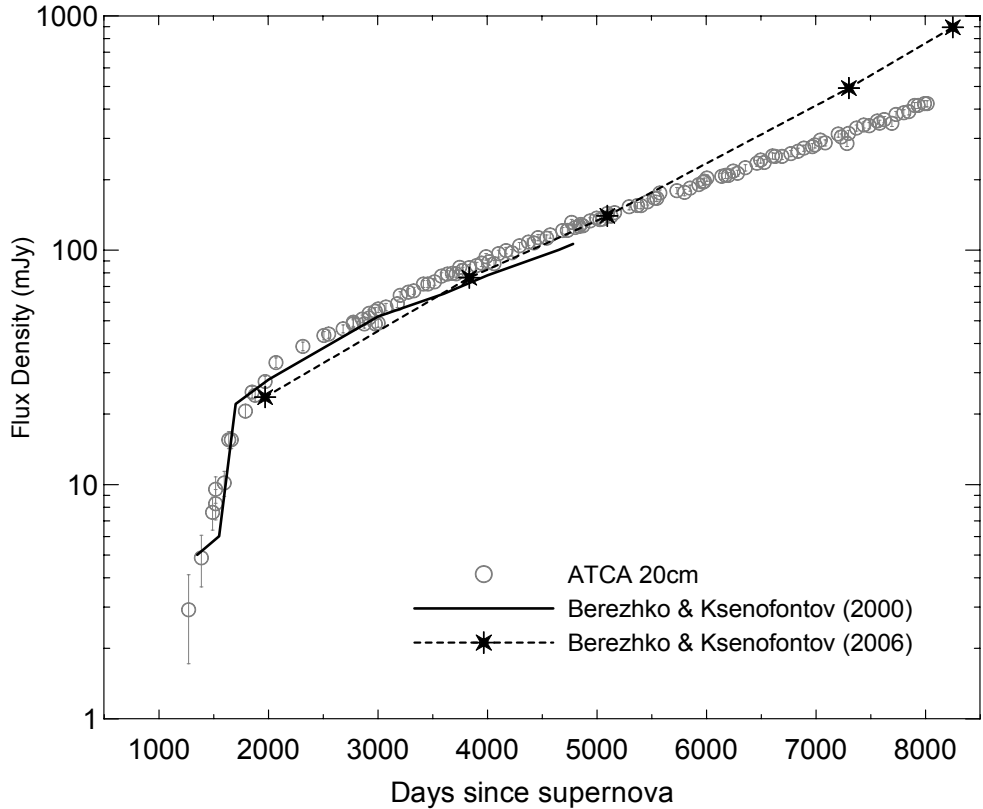}
\bigskip
\caption{Comparison of ATCA 1.4 GHz light curve (open gray circles) with models by Berezhko \& Ksenofontov. The continuous black line corresponds to the model based on the flux density data reported in Gaensler et al. (1997) from Day 918 to Day 3325, and provides theoretical predictions up to Day 4780 (Berezhko \& Ksenofontov 2000). The black asterisks are derived from the model presented in  Berezhko \& Ksenofontov (2006) that corresponds to the upstream magnetic field $B_{1}\sim$ 3 mG and $\sigma\sim 3$. This model  is calibrated on the radio data given by Manchester et al. (2002) for Days 1970, 3834 and 5093, and provides theoretical predictions for Day 7300 and Day 8249, which appear to overpredict the observed flux density by a factor of almost two.
\label{natural}}
\end{figure}
\clearpage

\begin{figure}[!ht]
\centering
\includegraphics[width=18 cm, angle=270]{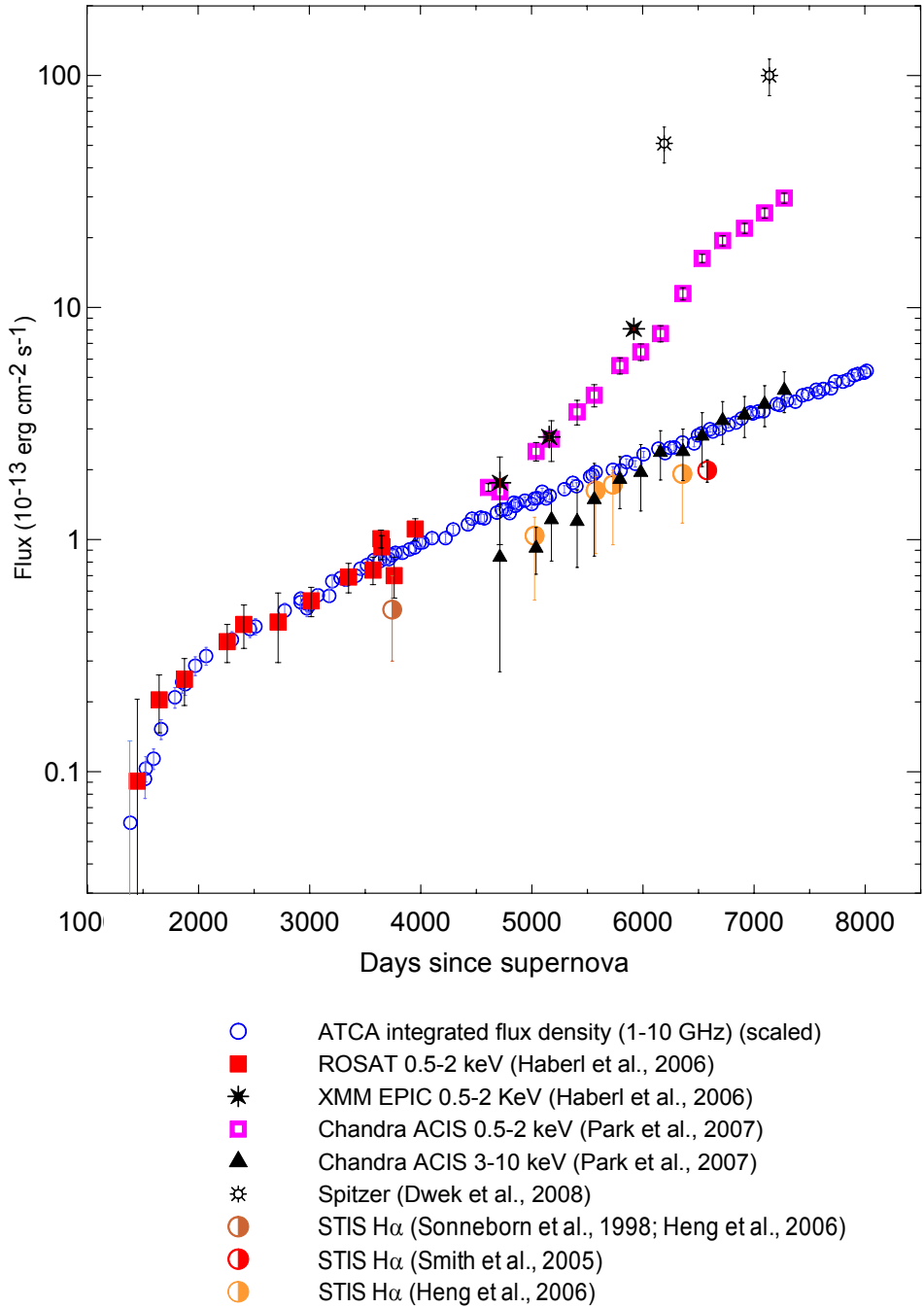}
\caption{Comparison of the radio flux with X-ray, H$\alpha$ and infrared fluxes. {\it Chandra} {\it ACIS} X-ray data (up to Day 7271)  are from Park at al. (2007); {\it ROSAT} data (up to Day 3950) and {\it XMM-Newton} data (Days 4712, 5156 and 5918) are from Haberl et al. (2006); H$\alpha$ flux values are from Sonneborn et al., (1998),  Smith et al. (2005) and Heng et  al. (2006); {\it Spitzer} data (Days 6190 and 7137) are from Dwek et al. (2008). The radio flux was derived after integrating the flux density in the 1-10 GHz frequency range. After being converted into X-ray flux units, radio flux values  have been multiplied by a factor of $2.7\times10^7$ to match the scale of {\it ROSAT} data.
\label{natural}}
\end{figure}

\begin{figure}[!ht]
\centering
\includegraphics[width=12.8 cm, angle=270]{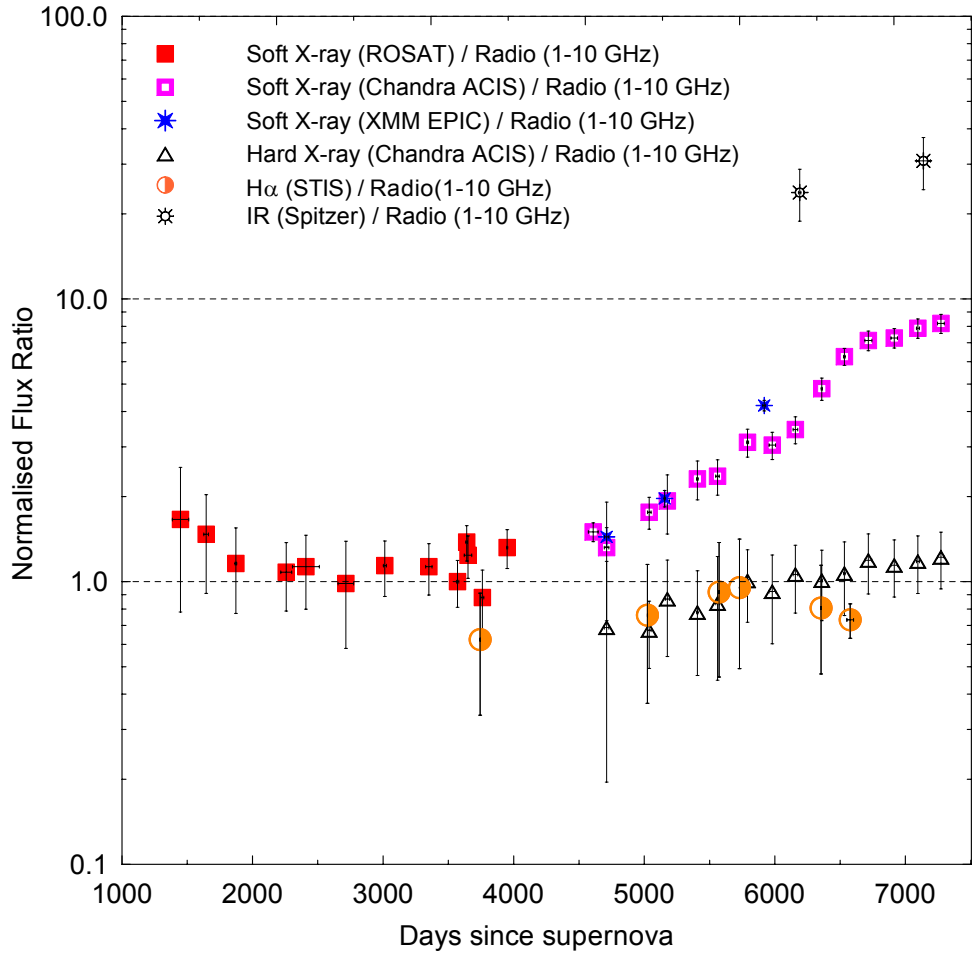}
\bigskip
\caption{Normalized ratio of the radio flux to the soft and hard X-ray, H$\alpha$ and IR fluxes. For data references see Figure 10.
\label{natural}}
\end{figure}

\clearpage

\end{document}